\documentclass[10pt]{article}
\usepackage{amsmath}
\usepackage{physics}
\usepackage{amsfonts}
\usepackage{amssymb}
\usepackage{graphicx}%
\usepackage{amstext}
\usepackage{latexsym}
\usepackage{color}
\usepackage{mathtools}
\usepackage{graphics}
\usepackage{cite}
\usepackage{slashed}
\usepackage{epsfig}
\usepackage{amsthm}
\usepackage{appendix}
\usepackage{hyperref}
\usepackage{multirow}
\usepackage{color}
\usepackage[normal]{subfigure}
\usepackage{rotating}
\usepackage{esdiff}
\usepackage[table]{xcolor}
\usepackage{enumitem}
\usepackage[utf8]{inputenc}
\usepackage{authblk}
\usepackage{colortbl}
\usepackage{pdflscape}
\usepackage{bm}
\usepackage{textcomp, gensymb}
\usepackage{gensymb}
\allowdisplaybreaks

\newcommand{\be}{\begin{equation}}
\newcommand{\en}{\end{equation}}

\newtheorem{defi}{Definition}[section]
\newtheorem{lem}[defi]{Lemma}

\newcommand{\bedefin}{\begin{defi}}
\newcommand{\findefi}{\end{defi} \medskip}

\newcommand{\belem}{\begin{lem}$\!\!${\bf }}
\newcommand{\enlem}{\end{lem}}

\newcommand{\beno}{\begin{equation*}}
\newcommand{\enno}{\end{equation*}}

\newcommand{\bea}{\begin{eqnarray}}
\newcommand{\ena}{\end{eqnarray}}

\catcode `\@=11 \@addtoreset{equation}{section}

\catcode `\@=12

\newcommand{\g}{G_{\hbox{\tiny{NC}}}}
\newcommand{\gd}{\hat{G}_{\hbox{\tiny{NC}}}}
\newcommand{\G}{\mathfrak{g}_{\hbox{\tiny{NC}}}}
\newcommand{\wh}{G_{\hbox{\tiny{WH}}}}

\newcommand{\anc}{\boldsymbol{A}^{\hbox{\tiny{nc}}}}

\begin{document}

\title{Supersymmetric Quantum Mechanics on a noncommutative plane through the lens of deformation quantization}

\author[1]{Md. Rafsanjany Jim\thanks{rafsanjany.jim@bracu.ac.bd}}
\author[1]{S. Hasibul Hassan Chowdhury\thanks{shhchowdhury@bracu.ac.bd}}

\affil[1]{Department of Mathematics and Natural Sciences, BRAC University, Kha 224 Bir Uttam Rafiqul Islam Avenue, Merul Badda, Dhaka, Bangladesh}

\maketitle

\begin{abstract}

A gauge invariant mathematical formalism based on deformation quantization is outlined to model an $\mathcal{N}=2$ supersymmetric system of a spin $1/2$ charged particle placed in a nocommutative plane under the influence of a vertical uniform magnetic field. The noncommutative involutive algebra $(C^{\infty}(\mathbb{R}^{2})[[\vartheta]],*^r)$ of formal power series in $\vartheta$ with coefficients in the commutative ring $C^{\infty}(\mathbb{R}^{2})$ was employed to construct the relevant observables, viz., SUSY Hamiltonian $H$, supercharge operator $Q$ and its adjoint $Q^{\dag}$ all belonging to the $2\times 2$ matrix algebra $\mathcal{M}_{2}(C^{\infty}(\mathbb{R}^{2})[[\vartheta]],*^r)$ with the help of a family of gauge-equivalent star products $*^{r}$. The energy eigenvalues of the SUSY Hamiltonian all turned out to be independent of not only the gauge parameter $r$ but also the noncommutativity parameter $\vartheta$. The nontrivial Fermionic ground state was subsequently computed associated with the zero energy which indicates that supersymmetry remains unbroken in all orders of $\vartheta$. The Witten index for the noncommutative SUSY Landau problem turns out to be $-1$ corroborating the fact that there is no broken supersymmetry for the model we are considering.

\end{abstract}

\section{Introduction}\label{sec:intro}

Supersymmetric quantum mechanics arises as the (0+1)-dimensional limit of the supersymmetric quantum field theory. E. Witten in \cite{Witten_constraints} used simple supersymmetric quantum systems to describe under what conditions supersymmetry is spontaneously broken. In \cite{Witten_supersymmetry_morse}, Witten carried on remarkably connecting the difference of the number of Bosonic ground states from that of the Fermionic ground states with a topological invariant intrinsic to the manifold associated with the supersymmetric quantum system. Under some 1-parameter deformation of conjugation type by a real valued function (Morse function) on the manifold, this topological invariant called Witten index stays unchanged. He used Morse theory there \cite{Witten_supersymmetry_morse} to establish this connection. Supersymmetric quantum mechanics in noncommutative spaces have been studied in the past (see, for example, \cite{Ashokdasetal}). The supersymmetric Landau problem in a noncommutative plane that we study in this article had been studied in the past in \cite{Harikumaretal, Halderetal,Ashokdasetal}. Although supersymmetry is shown to remain unbroken in \cite{Ashokdasetal}, the SUSY Hamiltonian there yields different spectra in the symmetric and Landau gauges as verified in appendix \ref{susec:appendix-ashokdasetal}. On the other hand, supersymmetry is shown to be broken in appendix \ref{app:Harikumar} for the cases of \cite{Harikumaretal,Halderetal}. All these are remedied in the present article by introducing a gauge invariant mathematical formalism for which supersymmetry remains unbroken in all order of the noncommutativity parameter $\vartheta$. 

The physical problem that we address in the present article is that of a spin $1/2$ charged particle of mass $m$ and charge $e$ constrained to move in a noncommutative plane under the influence of a uniform vertical magnetic field $B$. The dynamics of such a charged particle having spin in the commutative setup is best described by the Pauli Hamiltonian given by (\ref{eq:pauli_H}). Since the commutative problem exhibits supersymmetry (see \cite{Cooperetal}), we expect similar results to hold in the noncommutative setup. Essentially, there are 2 technical ingredients of the recipe: 1) Noncommutativity and 2) Supersymmetry. The first ingredient of the recipe for the model we are considering was introduced in \cite{Chowdhury1}. For the classical system of a particle moving in a 2-dimensional plane,  observables belong to the commutative algebra $C^{\infty}(\mathbb{R}^{2})$, equipped with pointwise product. Deformation quantization technique can be used by deforming this commutative algebra $C^{\infty}(\mathbb{R}^{2})$ of classical observables using a deformation parameter $\vartheta$ (for instance, by taking $F,G$ in (\ref{eq:star_product}) to be smooth functions on $\mathbb{R}^{2}$). For the quantum mechanical system of a scalar (spinless) particle moving in a 2-dimensional plane coupled to an external vertical uniform magnetic field, we need to consider equivalence classes of unitary irreducible representations (Unirreps) of an appropriate symmetry group. We know from our previous studies \cite{Chowdhury4} that the kinematical symmetry group, for a quantum mechanical system in a noncommutative plane, is the nilpotent Lie group $\g$ of the triple central extension of the abelian group of translations in $\mathbb{R}^{4}$. Therefore, we want our quantum observable space (which turns out to be an involutive noncommutative algebra after successful deformation of $C^{\infty}(\mathbb{R}^{2})$) to conform with a certain class of self adjoint irreducible representations (see \ref{eq:liegnc}) of the universal enveloping algebra $\mathcal{U}(\G)$ of the Lie algebra $\G$ associated with the Lie group $\g$ that models a family (parameterized by $r$) of equivalent noncommutative planes. But we have a particle on a noncommutative plane to be coupled with an external uniform vertical magnetic field. This is achieved by introducing the noncommutative $U(1)$ gauge field (\ref{eq:nc-gaugefields-exact-expression}). While applying the deformation quantization technique, we had to maintain careful agreement between the representation theoretic picture (see section \ref{sec:representation-theory-gnc}) and the deformation quantization picture (see section \ref{subsec:star-product}). Now that we have the mathematical framework to model a scalar charged particle in a noncommutative plane coupled to an external vertical uniform magnetic field using deformation quantization following \cite{Chowdhury1}, we proceed to formalise the framework to deal with the second ingredient of the recipe, namely supersymmetry.   

The quantum observables, in the mathematical formalism we propose, belong to the noncommutative $*$-algebra (or involutive algebra) $(C^{\infty}(\mathbb{R}^{2})[[\vartheta]],*^{r})$ of formal power series in $\vartheta$ with coefficients in the commutative ring $C^{\infty}(\mathbb{R}^{2})$ first introduced in \cite{Chowdhury1}. The noncommutative, associative product between 2 formal power series $F,G\in C^{\infty}(\mathbb{R}^{2})[[\vartheta]]$ is given in (\ref{eq:star_product}) which was first introduced in \cite{Chowdhury1}. The parameter $r$ that labels this noncommutative product is called the {\em gauge parameter} as the resulting Poisson structure is independent of it (see \ref{eq:Poisson-bracket-expression}). Note that the gauge parameter $r$ here that labels the star product $*^{r}$ is the same as the one that labels the equivalent noncommutative planes in (\ref{eq:liegnc}) as mentioned in the previous paragraph. The gauge parameter value $r=0$ corresponds to the familiar {\em Landau gauge} while $r=\frac{1}{2}$ corresponds to the {\em symmetric gauge}. The involution is defined in (\ref{eq:definition-involution}) using the inner product of the Hilbert space $L^{2}(\mathbb{R}^{2},\hbox{dx}\;\hbox{dy})$. The state space is a $\mathbb{Z}_{2}$-graded algebra $\mathcal{F}=((C^{\infty}(\mathbb{R}^{2})\cap L^{2}(\mathbb{R}^{2},\hbox{dx}\;\hbox{dy}))[[\vartheta]],*^r)\oplus ((C^{\infty}(\mathbb{R}^{2})\cap L^{2}(\mathbb{R}^{2},\hbox{dx}\;\hbox{dy}))[[\vartheta]],*^r)$ of formal power series in $\vartheta$. The first component $\psi^{(2)}$ of an element $\mathbf{\Psi}=\begin{pmatrix}\psi^{(2)}\\\psi^{(1)}\end{pmatrix}\in\mathcal{F}$ is meant to represent a bosonic state while the second component $\psi^{(1)}$ of $\mathbf{\Psi}$ represents the fermionic partner state of $\psi^{(2)}$. The super potential $\mathcal{A}\in(C^{\infty}(\mathbb{R}^{2})[[\vartheta]],*^r)$ is a formal power series in $\vartheta$ with coefficients in the commutative ring $C^{\infty}(\mathbb{R}^{2})$. The explicit expression of the superpotential $\mathcal{A}$ in terms of its $*^r$-product with a bosonic or a fermionic state $\psi\in((C^{\infty}(\mathbb{R}^{2})\cap L^{2}(\mathbb{R}^{2},\hbox{dx}\;\hbox{dy}))[[\vartheta]],*^r)$ is given in (\ref{eq:superpotential-action}) so that $\mathcal{A}*^{r}\psi\in((C^{\infty}(\mathbb{R}^{2})\cap L^{2}(\mathbb{R}^{2},\hbox{dx}\;\hbox{dy}))[[\vartheta]],*^r)$. The supercharge operator $Q$ and its adjoint $Q^{\dag}$ are $2\times 2$ matrices with entries in the noncommutative $*$-algebra $(C^{\infty}(\mathbb{R}^{2})[[\vartheta]],*^r)$, i.e., $Q,Q^{\dag}\in\mathcal{M}_{2}(C^{\infty}(\mathbb{R}^{2})[[\vartheta]],*^r)$ as given by (\ref{eq:nc_supercharges}). The supercharge operator $Q$ upon acting on a fermionic state $\begin{pmatrix}0\\\psi^{(1)}\end{pmatrix}\in\mathcal{F}$ yields a bosonic state $\begin{pmatrix}\mathcal{A}*^{r}\psi^{(1)}\\0\end{pmatrix}\in\mathcal{F}$ as can easily be seen using (\ref{eq:action-supercharge}). Along the same vein, the adjoint $Q^{\dag}$, of the supercharge operator $Q$, acting on a bosonic state $\begin{pmatrix}\psi^{(2)}\\0\end{pmatrix}\in\mathcal{F}$ returns the fermionic state $\begin{pmatrix}0\\\mathcal{A}^{\dag}*^{r}\psi^{(2)}\end{pmatrix}\in\mathcal{F}$ which also follows from (\ref{eq:action-supercharge}). One then writes down the supersymmetric (SUSY) Hamiltonian $H=\begin{pmatrix}H_{2}&0\\0&H_{1}\end{pmatrix}\in\mathcal{M}_{2}(C^{\infty}(\mathbb{R}^{2})[[\vartheta]],*^r)$, with $H_{2},H_{1}\in(C^{\infty}(\mathbb{R}^{2})[[\vartheta]],*^r)$ being the Bosonic and Fermionic Hamiltonians, respectively. The noncommutative eigenvalue equations for the partner Hamiltonians then read as
\begin{equation}\label{eq:eigenvalue-eq-partner-Hamiltonians}
\begin{aligned}
&H_{2}*^{r}\psi^{(2)}_{n}=E^{(2)}_{n}\psi^{(2)}_{n},\\
&H_{1}*^{r}\psi^{(1)}_{n+1}=E^{(1)}_{n+1}\psi^{(1)}_{n+1},
\end{aligned}
\end{equation}
for $n=0,1,2,...$. Energy eigenvaules $E^{(i)}_{n}$ of the partner Hamiltonians can be read off from the following expression
\begin{equation}\label{eq:explicit-eigenvalues-SUSY-Hamiltonian}
E^{(i)}_{n}=\hbar\frac{eB}{m}\left(n+\frac{1}{2}\right)\pm\frac{1}{2m}\hbar eB,
\end{equation}
with $n=0,1,2...$ above and one considers $+$ and $-$ on the right side of (\ref{eq:explicit-eigenvalues-SUSY-Hamiltonian}) for $i=2$ and $i=1$, respectively, on its left side. Here, $m$ is the mass and $e$ is the charge of the charged particle constrained to move on a $2$-dimensional noncommutative plane subjected to the uniform magnetic field $B$.

Equation (\ref{eq:eigenvalue-eq-partner-Hamiltonians}) then leads one to the eigenvalue equation of the SUSY Hamiltonian $H$ for the excited SUSY eigenstate $\mathbf{\Psi}_{n+1}=\begin{pmatrix}\psi^{(2)}_{n}\\\psi^{(1)}_{n+1}\end{pmatrix}\in\mathcal{F}$
\begin{equation}\label{eq:SUSY-eigenvalue-equation}
H\mathbf{\Psi}_{n+1}=E_{n+1}\mathbf{\Psi}_{n+1},
\end{equation}
with $n=0,1,2,....$ and the eigenvalue $E_{n+1}$ for the excited SUSY state $\mathbf{\Psi}_{n+1}$ is given by
\begin{equation}\label{eq:eigenvalues-SUSY_Hamiltonian}
E_{n+1}:=E_{n}^{(2)}=E_{n+1}^{(1)}=\hbar\frac{eB}{m}(n+1).
\end{equation}
Since $H\in\mathcal{M}_{2}(C^{\infty}(\mathbb{R}^{2})[[\vartheta]],*^r)$, on the left side of (\ref{eq:SUSY-eigenvalue-equation}), the entries, of the pertinent matrices following matrix multiplication, naturally get multiplied using the noncommutative $*^r$-product. One immediately finds using (\ref{eq:explicit-eigenvalues-SUSY-Hamiltonian}) that the ground state energy $E^{(1)}_{0}$ of the Fermionic Hamiltonian is zero. There is no nontrivial Bosonic eigenstate, i.e., eigenstate of $H_2$ with zero energy as is also evident from (\ref{eq:explicit-eigenvalues-SUSY-Hamiltonian}). Hence, the ground state $\mathbf{\Psi}_{0}$ of the SUSY Hamiltonian $H$ associated with zero eigenvalue can be read off as
\begin{equation}
\mathbf{\Psi}_{0}=\begin{pmatrix}0\\\psi^{(1)}_{0}\end{pmatrix},
\end{equation}
where the fermionic ground state $\psi^{(1)}_{0}$ is worked out in (\ref{eq:r-dependent-wave-function-main}) for an arbitrary real value of the gauge parameter $r$. It is worth remarking here that the Fermionic ground state (see \ref{eq:Landau-ground-state}) is independent of the noncommutativity parameter $\vartheta$ in the Landau gauge ($r=0$) while the Fermionic ground state (\ref{eq:symmetric-ground-state}) in the symmetric gauge (corresponding to $r=\frac{1}{2}$) explicitly depends on the noncommutativity parameter $\vartheta$. Similar situation arises when one computes the excited eigenstates of the SUSY Hamiltonian. The $n$\textsuperscript{th} eigenstate $\psi^{(1)}_{n,\hbox{\tiny{Lan}}}$ of the Fermionic Hamiltonian $H_1$ in the Landau gauge is computed in (\ref{eq:fermionic_excited_state_land}). This expression contains the $n$\textsuperscript{th} Hermite polynomial in the shifted $x$-coordinate exponentially suppressed by a factor containing the same shifting. The shifting does not involve any $\vartheta$ contribution. The superpartner of $\psi^{(1)}_{n,\hbox{\tiny{Lan}}}$, then, can easily be seen to be $\mathcal{A}*^{0}\psi^{(1)}_{n,\hbox{\tiny{Lan}}}$, which, in terms of superpotential operator (\ref{eq:ncsuperpotential}), can be read off as $\hat{\mathcal{A}}^{0}\psi^{(1)}_{n,\hbox{\tiny{Lan}}}$. One can then gather these excited partner states as components of a column vector (see \ref{eq:SUSY-eigenstate-Landau-gauge}) representing the $n$\textsuperscript{th} excited state of the SUSY Hamiltonian in the Landau gauge. The $n$\textsuperscript{th}  eigenstate $\psi^{(1)}_{n,\hbox{\tiny{Sym}}}$ (see \ref{eq:fermionic_excited_state_symmetric}) of the Fermionic Hamiltonian $H_1$ in the symmetric gauge, on the other hand, contains associated Laguerre polynomials in $\vartheta$ deformed radial variable in the cylindrical coordinate system. There are an exponentially suppressed and a positive exponential factors containing similar deformation of the radial coordinate of the underlying cylindrical system. One, then, finds the superpartner state $\mathcal{A}*^{\frac{1}{2}}\psi^{(1)}_{n,\hbox{\tiny{Sym}}}=\hat{\mathcal{A}}^{\frac{1}{2}}\psi^{(1)}_{n,\hbox{\tiny{Sym}}}$, of the Fermionic state $\psi^{(1)}_{n,\hbox{\tiny{Sym}}}$ which can be gathered together as components of a column vector (see \ref{eq:SUSY-eigenstate-symmetric-gauge}) representing the $n$\textsuperscript{th} excited state of the SUSY Hamiltonian in the symmetric gauge.

It is to be noted that in our formalism, the eigenvalues of the partner Hamiltonians in (\ref{eq:explicit-eigenvalues-SUSY-Hamiltonian}) and hence the eigenvalues of the SUSY Hamiltonian given by (\ref{eq:eigenvalues-SUSY_Hamiltonian}) are not only gauge independent (parameter $r$ independent) but also are all independent of the noncommutativity parameter $\vartheta$. The mathematical reasoning for $\vartheta$ independence of the SUSY energy spectra is that our formalism is a gauge independent one, i.e., one obtains the same energy (see \ref{eq:eigenvalues-SUSY_Hamiltonian}) for a given SUSY partner states (determined by $n\in\mathbb{N}$) in any gauge. Now, as emphasized in the previous paragraph, both the ground state $\psi^{(1)}_{0,\hbox{\tiny{Lan}}}$ and the $n$\textsuperscript{th} excited eigenstate $\psi^{(1)}_{n,\hbox{\tiny{Lan}}}$ of the Fermionic Hamiltonian $H_1$ in the Landau gauge does not contain $\vartheta$ in their expressions. As a result, there will be no contribution of $\vartheta$ in the pertaining energy eigenvalues of the Fermionic Hamiltonian in the Landau gauge. Since our formalism is a gauge independent one, energy eigenvalues of the SUSY Hamiltonian will be independent of $\vartheta$ in any gauge. In other words, energy spectra of the SUSY Hamiltonian for the SUSY Landau problem in a noncommutative plane exactly matches with the ones obtained from the SUSY Hamiltonian associated with the SUSY Landau problem in an ordinary 2-dimensional plane.

The organization of the article is as follows. In section \ref{sec:representation-theory-gnc}, we briefly recall the mathematical formalism adopted in \cite{Chowdhury1} for modelling a spinless charged particle moving on a noncommutative plane under the influence of an external vertical uniform magnetic field. Subsequently, in section \ref{sec:Supersymmetric Quantum Mechanics}, we extend the formalism to model a spin $\frac{1}{2}$ charged particle in the same set up  by embedding the formalism discussed in the previous section into the framework of $\mathcal{N}=2$ supersymmetry. In particular, we explicitly compute the superpotential as a formal power series in the deformation parameter $\vartheta$ using the noncommutative $U(1)$ gauge fields introduced in section \ref{subsec:star-product}. The pertinent super charge operator $Q$, its adjoint $Q^{\dag}$, and supersymmetric Hamiltonian $H$ were thereafter computed that satisfy the deformed SUSY algebra given by (\ref{eq:deformed-SUSY-algebra-N=2}). Gauge dependent ground state of the Fermionic Hamiltonian $H_1$ was explicitly computed there. We then obtained the $n$\textsuperscript{th} excited eigenstate of $H_1$ and its partner state explicitly. In section \ref{sec:comparison}, we compare the results obtained in this article with the ones obtained in some of the relevant published articles. Finally, in section \ref{sec:conclusion}, we provide our concluding remarks and point out some of the possible future directions. Appendices \ref{app:Harikumar},\ref{susec:appendix-ashokdasetal} were created to facilitate the comparisons conducted in section \ref{sec:comparison}, while appendix \ref{appendix:ground_state_of_fermionic_hamiltonian} provides the detailed calculation for the computation of the gauge parameter $r$-dependent ground state of the Fermionic Hamiltonian $H_1$.

%----- Charged Particle moving in A Noncommutative Plane and Minimal Coupling----%

\section{Scalar Charged Particle in a Noncommutative Plane and Minimal Coupling to an External Magnetic Field}\label{sec:representation-theory-gnc}

A triple central extension of the abelian group of translations in $\mathbb{R}^{4}$ was introduced in \cite{Chowdhury3, Chowdhury5}. The resulting centrally extended group is a 7-dimensional nilpotent Lie group which was denoted by $\g$ there. It is considered the kinematical symmetry group of quantum mechanics in a noncommutative plane (see \cite{Chowdhury4}) in the same spirit as the 5-dimensional  Heisenberg group $\wh$ underlies the structure of quantum mechanics in a 2-dimensional plane. For details on the comparison between $\g$ and $\wh$, refer to (section II of \cite{Chowdhury5}). The set of equivalence classes of unitary irreducible representations (unirreps) of $\g$ is called its unitary dual and is denoted by $\gd$. The unitary dual $\gd$ of $\g$ is computed in \cite{Chowdhury3} by means of the triple $(\hbar,\vartheta,B)$. Here, $B$ is interpreted as the magnetic field that is responsible for the noncommutativity between the 2 momentum operators out of the 4 noncentral generators of the nilpotent Lie group $\g$. In \cite{Chowdhury1}, one focuses on the equivalence class of unirreps of $\g$ labeled by $(\hbar,\vartheta,0)$. One then forms a family of equivalent irreducible self-adjoint representations (due to different values of the gauge parameter $r$) of the universal enveloping algebra $\mathcal{U}(\G)$ of the Lie algebra $\G$ associated with the Lie group $\g$:    

\begin{equation}\label{eq:liegnc}
    \begin{aligned}
        \hat{X}^r &= \hat{x}+\frac{(r-1)\vartheta}{\hbar}\hat{p}_y,\\
        \hat{Y}^r &= \hat{y}+\frac{r\vartheta}{\hbar}\hat{p}_x,\\
        \hat{\Pi}_x &= \hat{p}_x,\\
        \hat{\Pi}_y &= \hat{p}_y.
    \end{aligned}
\end{equation}
Here, $\hat{x}$ and $\hat{y}$ are the quantum mechanical position operators. They act on a generic element $\phi\in L^{2}(\mathbb{R}^{2},dx\;dy)$ in the following way
\begin{equation*}
\begin{aligned}
(\hat{x}\phi)(x,y)&=x\phi(x,y),\\
(\hat{y}\phi)(x,y)&=y\phi(x,y).
\end{aligned}
\end{equation*}
Also, the quantum mechanical momenta operators $\hat{p}_{x}$ and $\hat{p}_{y}$ act on a generic vector $\psi\in L^{2}(\mathbb{R}^{2},dx\;dy)$ in the following way
\begin{equation*}
\begin{aligned}
(\hat{p}_{x}\psi)(x,y)&=-i\hbar\frac{\partial\psi}{\partial x}(x,y),\\
(\hat{p}_{y}\psi)(x,y)&=-i\hbar\frac{\partial\psi}{\partial y}(x,y).
\end{aligned}
\end{equation*}

Physically, what one has is a family (parameterized by $r$) of noncommutative 2-planes due to a fixed triple $(\hbar,\vartheta,0)$. One then couples a spinless point particle of mass $m$, moving in the noncommutative plane, to an external uniform magnetic field $B$ minimally through the introduction of the following $1$-parameter family of representations ($e$ being the coupling):
\begin{equation}\label{eq:minimal-coupling-external-magnetic-field}
    \begin{aligned}
        \hat{X}^r &= \hat{x}+\frac{(r-1)\vartheta}{\hbar}\hat{p}_y,\\
        \hat{Y}^r &= \hat{y}+\frac{r\vartheta}{\hbar}\hat{p}_x,\\
        \hat{\Pi}^{r}_x &= \frac{2(1-r)e\hbar B}{\hbar+\sqrt{\hbar^2-4r(r-1)e\hbar\vartheta B}}\hat{y}+\Bigg[1+\frac{2r(1-r)e\vartheta B}{\hbar+\sqrt{\hbar^2-4r(r-1)e\hbar\vartheta B}}\Bigg]\hat{p}_x,\\
        \hat{\Pi}^{r}_y &= \frac{-2re\hbar B}{\hbar+\sqrt{\hbar^2-4r(r-1)e\hbar\vartheta B}}\hat{x}+ \Bigg[1+\frac{2r(1-r)e\vartheta B}{\hbar+\sqrt{\hbar^2-4r(r-1)e\hbar\vartheta B}}\Bigg]\hat{p}_y.
    \end{aligned}
\end{equation}

Here, $\hat{\Pi}^{r}_{x}$ and $\hat{\Pi}^{r}_{y}$ are called the kinematical momenta operators. The minimal coupling of the particle, of mass $m$ situated in the noncommutative plane (given by (\ref{eq:liegnc})), with the external electromagnetic field can be manifested by a simple manipulation of (\ref{eq:minimal-coupling-external-magnetic-field}):
\begin{equation}
    \begin{aligned}\label{eq:kinematical_momenta}
        \hat{X}^r &= \hat{x}+\frac{(r-1)\vartheta}{\hbar}\hat{p}_y,\\
        \hat{Y}^r &= \hat{y}+\frac{r\vartheta}{\hbar}\hat{p}_x,\\
        \hat{\Pi}^r_x &= \hat{p}_x+\frac{2e(1-r)\hbar B}{\hbar +\sqrt{\hbar^2-4r(r-1)e\hbar\vartheta B}}\Bigg(\hat{y}+\frac{r\vartheta}{\hbar}\hat{p}_x\Bigg),\\
        \hat{\Pi}^r_y &= \hat{p}_y - \frac{2re\hbar B}{\hbar +\sqrt{\hbar^2-4r(r-1)e\hbar\vartheta B}}\Bigg(\hat{x}+\frac{(r-1)\vartheta}{\hbar}\hat{p}_y\Bigg).
    \end{aligned}
\end{equation}

The representation given in the equation above satisfy the following set of commutation relations:
\begin{equation}\label{eq:liegnc-commutators}
    \begin{aligned}
        [\hat{X}^r,\hat{Y}^r] &= i\vartheta\hat{\mathbb{I}}, \quad [\hat{\Pi}^r_x,\hat{\Pi}^r_y]=ie\hbar B\hat{\mathbb{I}},\\
        [\hat{X}^r,\hat{\Pi}^r_x] &= \Bigg[1+\frac{2(1-r)e\vartheta B}{\hbar +\sqrt{\hbar^2-4r(r-1)e\hbar\vartheta B}}\Bigg]i\hbar \hat{\mathbb{I}},\\
        [\hat{Y}^r,\hat{\Pi}^r_y] &= \Bigg[1+\frac{2re\vartheta B}{\hbar +\sqrt{\hbar^2-4r(r-1)e\hbar\vartheta B}}\Bigg]i\hbar \hat{\mathbb{I}}.
    \end{aligned}
\end{equation}

Here, $\hat{\mathbb{I}}$ is the identity operator on $L^{2}(\mathbb{R}^{2},dx\;dy)$. The last 2 commutators $[\hat{X}^{r},\hat{\Pi}^{r}_{x}]$ and $[\hat{Y}^{r},\hat{\Pi}^{r}_{y}]$, above in (\ref{eq:liegnc-commutators}), are gauge dependent and are not intrinsic to the group $\g$ as the kinematical momenta $\hat{\Pi}^{r}_{x}$ and $\hat{\Pi}^{r}_{x}$ are not generators of the Lie group $\g$, but rather they are derived objects (see the last 2 equations of (\ref{eq:kinematical_momenta})) obtained from the group generators $\hat{X}^{r}$, $\hat{Y}^{r}$, $\hat{p}_{x}$ and $\hat{p}_{y}$ of the Lie group $\g$. Now, writing $\hat{\boldsymbol{\Pi}}^r \equiv (\hat{\Pi}^r_x,\hat{\Pi}^r_y)$ and $\hat{\boldsymbol{p}}\equiv (\hat{p}_x,\hat{p}_y)$, the kinematical momenta, in the last $2$ equations of (\ref{eq:kinematical_momenta}), take the following form
\begin{equation}
    \hat{\boldsymbol{\Pi}}^r = \hat{\boldsymbol{p}} -e\hat{\boldsymbol{A}^r},
\end{equation}
where the 2-component operator $\hat{\boldsymbol{A}}^r=(\hat{\boldsymbol{A}}^{r}_{x},\hat{\boldsymbol{A}}^{r}_{y})$ represents a noncommutative $\boldsymbol{U}(1)$ gauge field given by
\begin{equation}\label{eq:gauge-field-explicit}
    \hat{\boldsymbol{A}}^r = \Bigg(\frac{-2(1-r)\hbar B}{\hbar +\sqrt{\hbar^2-4r(r-1)e\hbar\vartheta B}}\hat{Y}^r,\frac{2r\hbar B}{\hbar +\sqrt{\hbar^2-4r(r-1)e\hbar\vartheta B}}\hat{X}^r\Bigg).
\end{equation}
The derivations $\partial_{i}\hat{\boldsymbol{A}}^{r}_{j}$, with $i,j=1,2$, of the components of the operatorial expression of the noncommutative $U(1)$ gauge field given by (\ref{eq:gauge-field-explicit}), are defined by 
\begin{equation}\label{eq:derivation-operatorial-gauge-fields}
\partial_{i}\hat{\boldsymbol{A}}^{r}_{j}=[\partial_{i},\hat{\boldsymbol{A}}^{r}_{j}],
\end{equation}
where $i,j=1,2$. From the explicit expression (\ref{eq:gauge-field-explicit}) of the operatorial representation of the 1-parameter family of noncommutative $U(1)$ gauge fields, one, using the definition (\ref{eq:derivation-operatorial-gauge-fields}) of derivation, obtains the following
\begin{equation}\label{eq:nonabelian-yang-mills-gauge-field}
\partial_{x}\hat{\boldsymbol{A}}^{r}_{y}-\partial_{y}\hat{\boldsymbol{A}}^{r}_{x}-\frac{ie}{\hbar}[\hat{\boldsymbol{A}}^{r}_{x},\hat{\boldsymbol{A}}^{r}_{y}]=B\hat{\mathbb{I}}.
\end{equation}
Equation (\ref{eq:gauge-field-explicit}) does not seem very natural as the underlying gauge fields are operator valued one forms. The more natural framework would be to treat the underlying  noncommutative U(1) gauge fields as some noncommutative algebra valued one forms. In the following subsection, we briefly discuss this framework called deformation quantization and discuss how it can be applied to the case of a scalar charged particle confined to a noncommutative plane under the influence of a vertical uniform magnetic field, as was carried out in \cite{Chowdhury1}.

\subsection{Star Product Approach}\label{subsec:star-product}
For a thorough understanding of the theory of deformation quantization from geometric point of view, we refer the readers to the beautiful exposition \cite{Fedosov} by Fedosov. In the context of a scalar charged particle moving in a noncommutative plane under the influence of a vertical uniform magnetic field $B$, one introduces a noncommutative associative *-algebra (involutive algebra) of formal power series in the deformation parameter $\vartheta$ (the same as the spatial noncommutativity parameter introduced in the previous section) and coefficients lying in the ring $C^{\infty}(\mathbb{R}^{2})$. 

To begin with, the set of all formal power series in $\vartheta$ with coefficients in the commutative ring $C^{\infty}(\mathbb{R}^{2})$ is endowed with the structure of a ring and is denoted by $C^{\infty}(\mathbb{R}^2)[[\vartheta]]$. Now, given 2 formal power series $F,G\in C^{\infty}(\mathbb{R}^{2})[[\vartheta]]$, one can define a noncommutative associative product between them following \cite{Chowdhury1} as follows     
\begin{equation}\label{eq:star_product}
    (F*^rG)(x,y)\coloneq F(x,y)e^{-i(r-1)\vartheta\overleftarrow{\partial_x}\overrightarrow{\partial_y}-ir\vartheta\overleftarrow{\partial_y}\overrightarrow{\partial_x}}G(x,y),
\end{equation}
which is parameterized by the same gauge parameter $r$, used to label the noncommutative planes in (\ref{eq:liegnc}). Here the formal power series $F,G$ can be read off as
\begin{equation}\label{eq:formal-power-series-expansion}
\begin{aligned}
F(x,y)&=\sum\limits_{n=0}^{\infty}F_{n}(x,y)\vartheta^{n},\\
G(x,y)&=\sum\limits_{n=0}^{\infty}G_{n}(x,y)\vartheta^{n},
\end{aligned}
\end{equation}
with $F_{n}(x,y)$ and $G_{n}(x,y)$ being smooth functions of $x,y$ for $n\geq 0$.

Endowed with the noncommutative associative $*^{r}$-product $*^r: C^{\infty}(\mathbb{R}^2)[[\vartheta]]\times C^{\infty}(\mathbb{R}^2)[[\vartheta]]\to C^{\infty}(\mathbb{R}^2)[[\vartheta]]$ defined in (\ref{eq:star_product}), the ring $C^{\infty}(\mathbb{R}^{2})[[\vartheta]]$ of formal power series now attains the structure of an involutive algebra $(C^{\infty}(\mathbb{R}^{2})[[\vartheta]],*^{r})$. The involution can easily be defined first by using the $*^{r}$-product of an element of $(C^{\infty}(\mathbb{R}^{2})[[\vartheta]],*^{r})$ with a sufficiently well-behaved function in $((C^\infty(\mathbb{R}^2)\cap L^2(\mathbb{R}^2,\hbox{dx}\,\hbox{dy}))[[\vartheta]],*^{r})$ and then using the inner product available in $L^{2}(\mathbb{R}^{2},\hbox{dx}\,\hbox{dy})$ (see (\ref{eq:definition-involution}) for details). The following result is a consequence of the star product defined above (\ref{eq:star_product}) 
\begin{equation}\label{eq:nc_xy}
    \begin{aligned}
        x*^r\psi &= \hat{X}^r\psi = x\psi-i(r-1)\vartheta\frac{\partial \psi}{\partial y},\\
        y*^r\psi &= \hat{Y}^r\psi = y\psi-ir\vartheta\frac{\partial \psi}{\partial x},
    \end{aligned}
\end{equation}
where $\psi$ is a sufficiently well-behaved vector in $((C^\infty(\mathbb{R}^2)\cap L^2(\mathbb{R}^2,\hbox{dx}\,\hbox{dy}))[[\vartheta]],*^{r})$. To uniformize notations, let us introduce the following definitions
\begin{equation}\label{eq:cc_pxpy}
    p_x *^r \psi \coloneq\hat{p}_x\psi = -i\hbar\frac{\partial \psi}{\partial x}, \qquad p_y*^{r}\psi \coloneq \hat{p}_y\psi=-i\hbar\frac{\partial \psi}{\partial y},
\end{equation}
where $p_x$ and $p_y$ can be considered as elements of the noncommutative $*$-algebra $(C^\infty(\mathbb{R}^2)[[\vartheta]],*^{r})$ and $\psi$ is a sufficiently well behaved functions in $((C^\infty(\mathbb{R}^2)\cap L^2(\mathbb{R}^2,\hbox{dx}\,\hbox{dy}))[[\vartheta]],*^{r})$.

Finally, one can proceed to show that the star product $*^r$ is gauge equivalent to the star product $*^{r^{\prime}}$ \cite{Chowdhury1} both obeying (\ref{eq:star_product}) by proving the existence of an invertible operator $T$ on $C^\infty(\mathbb{R}^2)$ satisfying
\begin{equation}
    T(F*^{r}G)=T(F)*^{r'}T(G).
\end{equation}
Here, the invertible operator is given by $T=e^{i(r-r')\vartheta \overrightarrow{\partial_x}\overrightarrow{\partial_y}}$.

According to this formalism, one can rewrite the vector potential given in operatorial form (see \ref{eq:gauge-field-explicit}) as
\begin{equation}\label{eq:nc-gaugefields-exact-expression}
    \anc \equiv (\anc_x,\anc_y)=\Bigg(\frac{-2(1-r)\hbar B}{\hbar +\sqrt{\hbar^2-4r(r-1)e\hbar\vartheta B}}y,\frac{2r\hbar B}{\hbar +\sqrt{\hbar^2-4r(r-1)e\hbar\vartheta B}}x\Bigg),
\end{equation}
where $x$ and $y$ are considered as elements of the noncommutative $*$-algebra $(C^\infty(\mathbb{R}^2)[[\vartheta]],*^{r})$ obeying(\ref{eq:nc_xy}). The gauge field written as an ordered pair of formal power series in $\vartheta$ now obeys the following in parallel with (\ref{eq:nonabelian-yang-mills-gauge-field}):
\begin{equation}\label{eq:nonabelian-yangmills-formal-powerseries}
\partial_x\anc_y-\partial_y\anc_x-\frac{ie}{\hbar}[\anc_x\!\overset{\;\;*^{r}}{,}\anc_y]=B\mathbb{I},
\end{equation}
where $\mathbb{I}\in C^{\infty}(\mathbb{R}^{2})$ that maps everything to $1$.

Before closing this section, let us look into how one obtains gauge parameter $r$-independent Poisson structure from the star product $*^r$ defined in (\ref{eq:star_product}). For $F,G\in(C^{\infty}(\mathbb{R}^{2})[[\vartheta]],*^r)$ as given by (\ref{eq:formal-power-series-expansion}), one obtains

\begin{equation}
    \begin{aligned}
        F*^rG&=F\exp\left[-i(r-1)\vartheta\overleftarrow{\partial_x}\overrightarrow{\partial_y}-ir\vartheta\overleftarrow{\partial_y}\overrightarrow{\partial_x}\right]G,\\
        &=F\left[1-i(r-1)\vartheta\overleftarrow{\partial_x}\overrightarrow{\partial_y}-ir\vartheta\overleftarrow{\partial_y}\overrightarrow{\partial_x}+\dots\right]G,\\
        &=FG-i(r-1)\vartheta\frac{\partial F}{\partial x}\frac{\partial G}{\partial y}-ir\vartheta\frac{\partial F}{\partial y}\frac{\partial G}{\partial x}+\dots.\\
    \end{aligned}
\end{equation}
Similarly,
\begin{equation}
    \begin{aligned}
        G*^rF&=G\exp\left[-i(r-1)\vartheta\overleftarrow{\partial_x}\overrightarrow{\partial_y}-ir\vartheta\overleftarrow{\partial_y}\overrightarrow{\partial_x}\right]F,\\
        &=G\left[1-i(r-1)\vartheta\overleftarrow{\partial_x}\overrightarrow{\partial_y}-ir\vartheta\overleftarrow{\partial_y}\overrightarrow{\partial_x}+\dots\right]F,\\
        &=FG-i(r-1)\vartheta\frac{\partial G}{\partial x}\frac{\partial F}{\partial y}-ir\vartheta\frac{\partial G}{\partial y}\frac{\partial F}{\partial x}+\dots.\\
    \end{aligned}
\end{equation}
One can, then, write the difference between the above two expressions of $*^{r}$-products as
\begin{equation}\label{eq:Poisson-structure-derivation}
    \begin{aligned}
        F*^rG-G*^rF&=-i(r-1)\vartheta\left[\frac{\partial F_0}{\partial x}\frac{\partial G_0}{\partial y}-\frac{\partial G_0}{\partial x}\frac{\partial F_0}{\partial y}\right]-ir\vartheta\left[\frac{\partial F_0}{\partial y}\frac{\partial G_0}{\partial x}-\frac{\partial G_0}{\partial y}\frac{\partial F_0}{\partial x}\right]+\dots\\
        &=i\vartheta\left[\frac{\partial F_0}{\partial x}\frac{\partial G_0}{\partial y}-\frac{\partial F_0}{\partial y}\frac{\partial G_0}{\partial x}\right]+\dots,
    \end{aligned}
\end{equation}
where the higher order terms in $\vartheta$ are dropped. Dividing both sides of (\ref{eq:Poisson-structure-derivation}) by $i\vartheta$ and taking the limit $\vartheta\rightarrow 0$, one obtains
\begin{equation}\label{eq:Poisson-bracket-expression}
\lim_{\vartheta\to 0}\frac{1}{i\vartheta}(F*^{r}G-G*^{r}F)=\{F_{0},G_{0}\},
\end{equation}
where the curly bracket on the right side of (\ref{eq:Poisson-bracket-expression}) stands for the Poisson bracket between the $0$-th order terms $F_{0}$ and $G_{0}$, of the formal power series $F$ and $G$ in $\vartheta$, respectively. We immediately see that although the star product between 2 formal power series involves the gauge parameter $r$ explicitly (see \ref{eq:star_product}), the term in their star commutator that is linear in $\vartheta$ (which is precisely the Poisson bracket between the pertaining 0-th order terms of the formal power series in question) is independent of the gauge parameter $r$, as expected. 
%---------------- Supersymmetric Quantum Mechanics ------------------%

\section{Supersymmetric Quantum Mechanics in Noncommutative Spaces}\label{sec:Supersymmetric Quantum Mechanics}
The main purpose of the present article is to study the system of a spin $\frac{1}{2}$ charged particle constrained to move on a 2-dimensional noncommutative plane under the influence of a uniform vertical magnetic field using the deformation quantization technique discussed in the previous section for the case of a scalar charged particle in a noncommutative plane subjected to a uniform vertical magnetic field. In the commutative setup, one can extend the formalism for the spinless charged particle to the case of a charged particle with spin by taking the coupling of the spin of the particle with the applied magnetic field into account. For the dynamics of a particle having spin there will be an extra contribution to the Hamiltonian due to this direct interaction of the intrinsic magnetic moment of the particle with the applied magnetic field (see page 456, \cite{Landau}). One includes this extra contribution to the Hamiltonian of a spinless charged particle to write the correct expression of the Hamiltonian of a charged particle having spin. A well-known Hamiltonian, that incorporates all these ideas for a particle with spin, is the Pauli Hamiltonian. In the following, we will first discuss the Pauli Hamiltonian in the commutative setup in detail. 

The system of a spin $\frac{1}{2}$ charged particle moving in an ordinary 2-dimensional plane under the influence of a vertical uniform magnetic field is known to exhibit supersymmetry \cite{Cooperetal}. 
The Hamiltonian corresponding to this commutative problem is given by the following Pauli Hamiltonian

\begin{equation}\label{eq:pauli_H}
    \hat{H}=\frac{1}{2m}\Big[(\hat{p}_i-e\hat{A}_i)^2\hat{\mathbb{I}}_{2\times 2}+\frac{g}{2}\hbar eB\hat{\sigma}_{3}\Big],
\end{equation}
where $m, e$ are the corresponding mass, and magnitude of the charge (hence nonnegative) of the particle, respectively, and $B$ is the magnitude of the magnetic field (hence nonnegative).  Besides, $\hat{\mathbb{I}}_{2\times 2}$ and $\hat{\sigma}_3$ 
are the $(2\times 2)$ identity matrix and the third Pauli matrix, respectively. The value of the 
gyro-magnetic ratio $g$ is 2 for the problem to possess supersymmetry. It is worth remarking here that supersymmetry (SUSY) fixes the value of the gyro-magnetic ratio $g$.  

The pertinent supercharge operator for the above defined Hamiltonian is given by
\begin{equation}\label{eq:c_supercharg_matrix}
    \hat{Q}=\frac{1}{\sqrt{2m}}
    \begin{pmatrix}
        0 & \hat{\mathcal{A}}\\
        0 & 0
    \end{pmatrix},
\end{equation}
where the superpotential $\hat{\mathcal{A}}$ operator is defined by
\begin{equation}\label{eq:s_potential}
    \hat{\mathcal{A}}=i(\hat{p}_x-e\hat{A}_x)+(-\hat{p}_y+e\hat{A}_y).
\end{equation}

Furthermore, the operators $\hat{Q},\hat{Q}^{\dag}$ and $\hat{H}$ satisfy the supersymmetric algebra, i.e.,
\begin{equation}
    \{\hat{Q},\hat{Q}\}=\{\hat{Q}^{\dag},\hat{Q}^{\dag}\}=0,\quad \{\hat{Q},\hat{Q}^{\dag}\}=\hat{H},\quad [\hat{H},\hat{Q}]=[\hat{H},\hat{Q}^{\dag}]=0,
\end{equation}
with $H$ is given in (\ref{eq:pauli_H}). One should also note that $\hat{H}$ is a $(2\times 2)$-operator valued matrix given by
\begin{equation}
    \hat{H}=\begin{pmatrix}
        \hat{H}_2 & 0\\
        0 & \hat{H}_1
    \end{pmatrix},
\end{equation}
where 
\begin{equation}
    \begin{aligned}
        \hat{H}_2&=\frac{1}{2m}\hat{\mathcal{A}}\hat{\mathcal{A}}^\dag,\\
        \hat{H}_1&=\frac{1}{2m}\hat{\mathcal{A}}^\dag\hat{\mathcal{A}},
    \end{aligned}
\end{equation}
are the partner Hamiltonians.

Henceforth, we will distinguish an operator $\hat{\mathcal{A}}$ on $L^{2}(\mathbb{R}^{2}, \hbox{dx}\;\hbox{dy})$ from the corresponding noncommutative algebra element $\mathcal{A}$ in $(C^{\infty}(\mathbb{R}^{2})[[\vartheta]],*^{r})$ with the introduction of a hat. It should also be mentioned here that in the above we consider $\hat{Q}$, $\hat{Q}^\dag$, and $\hat{H}$ to be operator (on $L^{2}(\mathbb{R}^{2},\hbox{dx}\,\hbox{dy})$)-valued $(2\times 2)$ matrices, while, in the following, we will construct analogous objects, denoted by $Q$, $Q^\dag$, and $H$, which are all $(2\times 2)$ noncommutative algebra $(C^{\infty}(\mathbb{R}^{2})[[\vartheta]],*^{r})$-valued matrices, i.e., $Q, Q^{\dag},H\in\mathcal{M}_{2}(C^{\infty}(\mathbb{R}^{2})[[\vartheta]],*^{r})$.

In this article, we consider the noncommutative $U(1)$ gauge field as an ordered pair of formal power series each belonging to $(C^\infty(\mathbb{R}^2)[[\vartheta]],*^{r})$ (see \cite{Chowdhury1})
\begin{equation}
    \boldsymbol{A}^{\hbox{\tiny{nc}}}\equiv (\boldsymbol{A}^{\hbox{\tiny{nc}}}_x,\boldsymbol{A}^{\hbox{\tiny{nc}}}_y)=\Bigg(\frac{-2(1-r)\hbar B}{\hbar +\sqrt{\hbar^2-4r(r-1)e\hbar\vartheta B}}y,\frac{2r\hbar B}{\hbar +\sqrt{\hbar^2-4r(r-1)e\hbar\vartheta B}}x\Bigg).
\end{equation}
 
The superpotential in terms of the formal power series $\boldsymbol{A}^{\hbox{\tiny{nc}}}_x$ and $\boldsymbol{A}^{\hbox{\tiny{nc}}}_y$ takes the following form
\begin{equation}\label{eq:ncsuperpotential}
    \mathcal{A}= i\Bigg(p_x-\frac{-2(1-r)e\hbar B}{\hbar +\sqrt{\hbar^2-4r(r-1)e\hbar\vartheta B}}y\Bigg)+\Bigg(-p_y+\frac{2re\hbar B}{\hbar +\sqrt{\hbar^2-4r(r-1)e\hbar\vartheta B}}x\Bigg).
\end{equation}
It is evident from equation (\ref{eq:ncsuperpotential}) that the superpotential is again in $(C^\infty(\mathbb{R}^2)[[\vartheta]],*^{r})$. The $*^r$-product between the superpotential $\mathcal{A}$ given by (\ref{eq:ncsuperpotential}) and a sufficiently well-behaved vector $\psi\in((C^\infty(\mathbb{R}^2)\cap L^2(\mathbb{R}^2,\hbox{dx}\,\hbox{dy}))[[\vartheta]],*^{r})$ is given by
\begin{equation}\label{eq:super_potential_product}
    \begin{aligned}
        \mathcal{A}*^r \psi&= i\Bigg[p_x*\psi-\frac{-2(1-r)e\hbar B}{\hbar +\sqrt{\hbar^2-4r(r-1)e\hbar\vartheta B}}y*^r\psi\Bigg]\\
        &\quad+\Bigg[-p_y*\psi+\frac{2re\hbar B}{\hbar +\sqrt{\hbar^2-4r(r-1)e\hbar\vartheta B}}x*^r\psi\Bigg],\\
        &= i\Bigg[\hat{p}_x\psi-\frac{-2(1-r)e\hbar B}{\hbar +\sqrt{\hbar^2-4r(r-1)e\hbar\vartheta B}}\hat{Y}^r\psi\Bigg]\\
        &\quad+\Bigg[-\hat{p}_y\psi+\frac{2re\hbar B}{\hbar +\sqrt{\hbar^2-4r(r-1)e\hbar\vartheta B}}\hat{X}^r\psi\Bigg].
    \end{aligned}
\end{equation}
Here, $\hat{X}^r$ and $\hat{Y}^r$ are given by the equation (\ref{eq:liegnc}). The above equation (\ref{eq:super_potential_product}) can be expressed as a formal power series in $\vartheta$ with coefficients in $C^{\infty}(\mathbb{R}^{2})\cap L^{2}(\mathbb{R}^{2},\hbox{dx}\,\hbox{dy})$:
\begin{equation}\label{eq:superpotential-action}
    \begin{aligned}
        \mathcal{A}*^r \psi&= \Big(\hbar\frac{\partial \psi}{\partial x}+i(1-r)eBy\psi-i\hbar\frac{\partial \psi}{\partial y}+reBx\psi\Big)\\
        &-\Bigg[i(r-1)^2eB\frac{\partial \psi}{\partial y}+\frac{ir(r-1)e^2B^2}{\hbar}y\psi+ir(r-1)eB\frac{\partial \psi}{\partial y}\\
        &-\frac{r^2(r-1)^2e^2B^2}{\hbar}x\psi\Bigg]\vartheta+\Bigg[\frac{2ir^2(r-1)e^3B^3}{\hbar^2}y\psi\\
        &-\frac{r(r-1)^2e^2B^2}{\hbar}\frac{\partial \psi}{\partial y}-\frac{ir^2(r-1)^3e^2B^2}{\hbar}\frac{\partial \psi}{\partial y}\\
        &+\frac{2r^3(r-1)^2e^3B^3}{\hbar^2}x\psi\Bigg]\vartheta^2+\mathcal{O}(\vartheta^3).
    \end{aligned}
\end{equation}
We denote this fact as $\mathcal{A}*^{r}\psi\in ((C^{\infty}(\mathbb{R}^{2})\cap L^{2}(\mathbb{R}^{2},\hbox{dx}\,\hbox{dy})[[\vartheta]],*^r)$, the noncommutative, associative $*$-algebra of formal power series in $\vartheta$ with coefficients in $C^{\infty}(\mathbb{R}^{2})\cap L^{2}(\mathbb{R}^{2},\hbox{dx}\,\hbox{dy})$. Therefore, one can write $Q$ and $Q^{\dag}$ in terms of $\mathcal{A}$. The Hamiltonian is the anticommutator of the supercharges $Q$ and $Q^{\dag}$. In particular, the supercharges $Q$ and $Q^{\dag}$ in the noncommutative plane is given by
\begin{equation}\label{eq:nc_supercharges}
    Q=\frac{1}{\sqrt{2m}}\begin{pmatrix}
        0 & \mathcal{A}\\
        0 & 0
    \end{pmatrix},\qquad
    Q^{\dag}=\frac{1}{\sqrt{2m}}\begin{pmatrix}
        0 & 0\\
        \mathcal{A}^{\dag} & 0
    \end{pmatrix}.
\end{equation}
Here, $Q$ and $Q^\dag$ are $(C^\infty(\mathbb{R}^2)[[\vartheta]],*^r)$-valued $2\times 2$ matrices, i.e., $Q,Q^\dag \in \mathcal{M}_2(C^\infty(\mathbb{R}^2)[[\vartheta]],*^r)$. One should also note here that we denote by $\dag$ the involution operation on the noncommutative $*$-algebra $(C^\infty(\mathbb{R}^2)[[\vartheta]],*^r)$. The involution $\mathcal{A}^{\dag}$ of the noncommutative algebra element $\mathcal{A}\in(C^\infty(\mathbb{R}^2)[[\vartheta]],*^r)$ can be realized by means of the $*^r$-product with a sufficiently well-behaved algebra element $\psi\in((C^\infty(\mathbb{R}^2)\cap L^2(\mathbb{R}^2,\hbox{dx}\,\hbox{dy}))[[\vartheta]],*^{r})$:
\begin{equation}\label{eq:definition-involution}
    \begin{aligned}
        \langle \mathcal{A}^{\dag}*^{r}\psi,\phi\rangle &=\langle(\hat{\mathcal{A}}^r)^{\dag}\psi,\phi\rangle,\\
        &=\langle \psi,\hat{\mathcal{A}}^{r}\phi\rangle,\\
        &=\langle\psi,\mathcal{A}*^{r}\phi\rangle,
    \end{aligned}
\end{equation}
where $\phi,\psi\in ((C^\infty(\mathbb{R}^2)\cap L^2(\mathbb{R}^2,\,dx\,dy))[[\vartheta]],*^r)$, $\langle\;,\;\rangle$ denotes the inner product of $L^2(\mathbb{R}^2,\,dx\,dy)$. One should also note here that $\mathcal{A}$ is a formal power series in $(C^\infty(\mathbb{R}^2)[[\vartheta]],*^r)$, whereas $\hat{\mathcal{A}}^{r}$ is an operator.

In supersymmetry, one considers a $\mathbb{Z}_2$-graded Hilbert space on which the supercharges act. In the context of deformation quantization, one can replace this $\mathbb{Z}_2$-graded Hilbert space by a $\mathbb{Z}_2$ graded algebra of formal power series:
\begin{equation}\label{eq:Z_2_graded_vector_space}
    \mathcal{F}=((C^{\infty}(\mathbb{R}^{2})\cap L^{2}(\mathbb{R}^{2},\hbox{dx}\;\hbox{dy}))[[\vartheta]],*^r)\oplus ((C^{\infty}(\mathbb{R}^{2})\cap L^{2}(\mathbb{R}^{2},\hbox{dx}\;\hbox{dy}))[[\vartheta]],*^r).
\end{equation}
Therefore, each element $\boldsymbol{\Psi}\in\mathcal{F}$ is a two component column vector, i.e.,
\begin{equation}
    \boldsymbol{\Psi}=\begin{pmatrix}
        \psi^{(2)}\\
        \psi^{(1)}
    \end{pmatrix},
\end{equation}
where each of $\psi^{(2)}$ and $\psi^{(1)}$ is in $((C^{\infty}(\mathbb{R}^{2})\cap L^{2}(\mathbb{R}^{2},\hbox{dx}\;\hbox{dy}))[[\vartheta]],*^r)$. As a result, the action of $Q$ (or $Q^\dag$) on $\boldsymbol{\Psi}\in \mathcal{F}$ is given by
\begin{equation}\label{eq:action-supercharge}
    \begin{aligned}
        Q\boldsymbol{\Psi} &= \frac{1}{\sqrt{2m}}\begin{pmatrix}
        0 & \mathcal{A}\\
        0 & 0
    \end{pmatrix}
    \begin{pmatrix}
        \psi^{(2)}\\
        \psi^{(1)}
    \end{pmatrix}=\frac{1}{\sqrt{2m}}\begin{pmatrix}
        \mathcal{A}*^r \psi^{(1)}\\
        0
    \end{pmatrix},\\
    Q^\dag \boldsymbol{\Psi} &=\frac{1}{\sqrt{2m}}\begin{pmatrix}
        0 & 0\\
        \mathcal{A}^\dag & 0
    \end{pmatrix}\begin{pmatrix}
        \psi^{(2)}\\
        \psi^{(1)}
    \end{pmatrix}=\frac{1}{\sqrt{2m}}\begin{pmatrix}
        0\\
        \mathcal{A}^\dag*^r\psi^{(2)}
    \end{pmatrix}.
    \end{aligned}
\end{equation}
The product defined above between $Q$ and $\boldsymbol{\Psi}$ merits elaboration: initially the product between $Q$ and $\boldsymbol{\Psi}$ follow the rule of matrix product while the product between the elements are the star product $*^r$.

Now,  one can find the partner Hamiltonians by straightforward calculations:
\begin{equation}\label{eq:Bosonic-Hamiltonian}
    \begin{aligned}
        H_2 &= \frac{1}{2m}\Big[(ip_x-ie\anc_x-p_y+e\anc_y)\ast^r(-ip_x+ie\anc_x-p_y+e\anc_y)\Big]\\
        &= \frac{1}{2m}\Big[p_x \ast^r p_x -ep_x\ast^r \anc_x-ip_x \ast^r p_y+iep_x\ast^r \anc_y-e\anc_x\ast^r p_x\\
        &\,+\, e^2\anc_x \ast^r \anc_x+ie\anc_x\ast^r p_y-ie^2\anc_x\ast^r \anc_y + ip_y\ast^r p_x-iep_y\ast^r \anc_x\\
        &\,+\, p_y\ast^r p_y - ep_y\ast^r \anc_y -ie\anc_y\ast^r p_x + ie^2\anc_y \ast^r \anc_x - e\anc_y\ast^r p_y +e^2\anc_y\ast^r \anc_y\Big],\\
        &= \frac{1}{2m}\Big[p_x \ast^r p_x - ep_x \ast^r \anc_x - e\anc_x\ast^r p_x +e^2\anc_x\ast^r\anc_x + p_y \ast^r p_y - e p_y \ast^r \anc_y \\
        & \,+\, e \anc_y\ast^r p_y +e^2\anc_y \ast^r \anc_y +ie\anc_x\ast^r p_y -ie \anc_y \ast^r p_x \\
        &\,+\, ie(p_x \ast^r \anc_y - p_y\ast^r \anc_x - e[\anc_x \!\overset{\;\;*^{r}}{,}\anc_y])\Big].
    \end{aligned}
\end{equation}
Similarly,
\begin{equation}\label{eq:fermionic_hamiltonian}
    \begin{aligned}
        H_1 &= \frac{1}{2m}\Big[p_x *^r p_x -ep_x*^r \anc_x -e\anc_x *^r p_x+e^2\anc_x *^r \anc_x+p_y*^rp_y-ep_y*^r\anc_y\\
        &\,-\, e\anc_y*^r p_y + e^2\anc_y*^r\anc_y-ie\anc_x*^rp_y+ie\anc_y*^rp_x \\
        &\,-\, ie(p_x*^r\anc_y-p_y*^r\anc_x-e[\anc_x\!\overset{\;\;*^{r}}{,}\anc_y])\Big].
    \end{aligned}
\end{equation}
Here, the part inside the braces of the partner Hamiltonians can be replaced by the following equation \cite{Chowdhury1}
\begin{equation}\label{eq:sw_map}
    \partial_x\anc_y-\partial_y\anc_x-\frac{ie}{\hbar}[\anc_x\!\overset{\;\;*^{r}}{,}\anc_y]=B\mathbb{I},
\end{equation}
where $B$ is the noncommutative field strength and $\mathbb{I}$ is the constant function in $C^\infty(\mathbb{R}^2)$ that maps all of $\mathbb{R}^{2}$ to the constant real value $1$. Using (\ref{eq:sw_map}), 
one can rewrite the partner Hamiltonians as
\begin{equation}\label{eq:partner_hamiltonian}
    \begin{aligned}
        H_2&=\frac{1}{2m}\Big[(p_x-e\anc_x)*^r(p_x-e\anc_x)+(p_y-e\anc_y)*^r(p_y-e\anc_y)+e\hbar B\mathbb{I}\Big],\\
        H_1&=\frac{1}{2m}\Big[(p_x-e\anc_x)*^r(p_x-e\anc_x)+(p_y-e\anc_y)*^r(p_y-e\anc_y)-e\hbar B\mathbb{I}\Big].
    \end{aligned}
\end{equation}
Therefore, the supersymmetric Hamiltonian is exactly the Pauli Hamiltonian, i.e.,
\begin{equation}\label{eq:SUSY-Hamiltonian-final-form}
    \begin{aligned}
        H=\frac{1}{2m}\Big[\Big((p_x-e\anc_x)*^r(p_x-e\anc_x)+(p_y-e\anc_y)*^r(p_y-e\anc_y)\Big)\hat{\mathbb{I}}_{2\times 2}+e\hbar B\mathbb{I}\hat{\sigma}_3\Big].
    \end{aligned}
\end{equation}
One should immediately notice from the above equation that the SUSY Hamiltonian is a $2\times 2$ matrix given by
\begin{equation}
    H=\begin{pmatrix}
            H_2 & 0\\
            0 & H_1
        \end{pmatrix}.
\end{equation}
Since $H_{2},H_{1}\in(C^{\infty}(\mathbb{R}^{2})[[\vartheta]],*^{r})$ as is evident from (\ref{eq:partner_hamiltonian}), one finds that $H\in\mathcal{M}_{2}(C^{\infty}(\mathbb{R}^{2})[[\vartheta]],*^{r})$.

Without much difficulties, one can now show that $Q,Q^{\dag}$ and $H$ satisfiy the following deformed SUSY algebra
\begin{equation}\label{eq:deformed-SUSY-algebra-N=2}
    \{Q,Q\}=\{Q^\dag , Q^\dag\}=0, \quad \{Q,Q^\dag\}=H,\quad [H,Q]=[H,Q^\dag]=0.
\end{equation}
Here, the product between any two matrices from $Q,Q^\dag$, and $H$ involved in the commutator and the anticommutator above in (\ref{eq:deformed-SUSY-algebra-N=2}) is just the matrix product while the product between elements of the relevant matrices ensuing from the matrix product is given by  the star product $*^r$. 

The {\em Witten operator} of the $\mathbb{Z}_{2}$-graded vector space $\mathcal{F}$, given by the equation (\ref{eq:Z_2_graded_vector_space}), can be read off as
\begin{equation}
    (-1)^F=\begin{pmatrix}
        \mathbb{I} & 0\\
        0 & -\mathbb{I}
    \end{pmatrix},
\end{equation}
where $\mathbb{I}$, being the constant function in $C^{\infty}(\mathbb{R}^{2})$ introduced earlier, is also the unital element of the noncommutative algebra $(C^\infty(\mathbb{R}^2)[[\vartheta]],*^r)$. One, therefore, has $(-1)^{F}\in\mathcal{M}_{2}(C^{\infty}(\mathbb{R}^{2})[[\vartheta]],*^{r})$. The eigenvalues of the Witten operator are $\pm 1$, as expected. Here, $1$ is the eigenvalue of $(-1)^F$ when it acts on a purely bosonic state of the form $\begin{pmatrix}
\psi^{(2)}\\0\end{pmatrix}\in\mathcal{F}$, while, $-1$ is the eigenvalue of  $(-1)^F$ when it acts on a Fermionic state $\begin{pmatrix} 0\\\psi^{(1)}\end{pmatrix}\in\mathcal{F}$. Furthermore, one can show that the Witten operator $(-1)^F$ satisfies the following equations involving commutator and anticommutator with $H$, $Q$, $Q^{\dag}$, and $(-1)^{F}$ itself (the matrix product is respected below again with the entries of the relevant matrices multiplied with one another using $*^{r}$ product):
\begin{equation}\label{eq:paraity-commutation-relation}
    [(-1)^F,H]=\boldsymbol{0}_{2\times 2}, \;\; \{(-1)^F, Q\}=\{(-1)^F,Q^\dag\}=\boldsymbol{0}_{2\times 2},\;\;\left((-1)^F\right)^2=\begin{pmatrix}\mathbb{I}&0\\0&\mathbb{I}\end{pmatrix}=:\mathbb{I}_{2\times2},
\end{equation}
where $\boldsymbol{0}_{2\times 2}\in\mathcal{M}_{2}(C^{\infty}(\mathbb{R}^{2})[[\vartheta]],*^{r})$ is the $2\times 2$ matrix with every element of it being the constatnt smooth $0$-function on $\mathbb{R}^{2}$ that maps all of $\mathbb{R}^{2}$ to the constant real value $0$. Also, $\mathbb{I}_{2\times 2}$ belongs to $\mathcal{M}_{2}(C^{\infty}(\mathbb{R}^{2})[[\vartheta]],*^{r})$. Therefore, the SUSY quantum mechanical system, corresponding to the model we are considering, can be represented by the following quintuple (consult \cite{Witten_supersymmetry_morse} for details)

\begin{equation}
    \left(H, Q, Q^\dag, (-1)^F, \mathcal{F}\right).
\end{equation}

From (\ref{eq:partner_hamiltonian}) and (\ref{eq:SUSY-Hamiltonian-final-form}), it is immediate that both the partner Hamiltonians, i.e. the Fermionic Hamiltonian $H_1$ and the Bosonic Hamiltonian $H_2$, consist of two parts. The first part
\begin{equation}
    \frac{1}{2m}\left[(p_x-e\boldsymbol{A}_x^{\hbox{\tiny nc}})*^r(p_x-e\boldsymbol{A}_x^{\hbox{\tiny nc}})+(p_y-e\boldsymbol{A}_y^{\hbox{\tiny nc}})*^r(p_y-e\boldsymbol{A}_y^{\hbox{\tiny nc}})\right]
\end{equation}
is exactly the Hamiltonian of the Landau problem in a noncommutative plane and the second part
\begin{equation}
    \frac{1}{2m}e\hbar B\mathbb{I}
\end{equation}
is an extra contribution. Using the technique specified in the reference (page 10 in \cite{Chowdhury1}), one can immediately write the n\textsuperscript{th} energy eigenvalue of the partner Hamiltonians $H_1$ and $H_2$ as
\begin{equation}
    \begin{aligned}
        E_n^{(1)}&=\hbar\frac{eB}{m}\left(n+\frac{1}{2}\right)-\frac{1}{2m}\hbar eB,\\
        E_n^{(2)}&=\hbar\frac{eB}{m}\left(n+\frac{1}{2}\right)+\frac{1}{2m}\hbar eB,
    \end{aligned}
\end{equation}
for $n\geq 0$. The above expression can also be written in a matrix form as
\begin{equation}\label{eq:eigenvalue_susy}
    \begin{pmatrix}
        E_n^{(2)} & 0\\
        0 & E_n^{(1)}
    \end{pmatrix}=\hbar\frac{eB}{m}\left(n+\frac{1}{2}\right)\hat{\mathbb{I}}_{2\times 2}+\frac{1}{2m}\hbar eB\hat{\sigma}_3,
\end{equation}
where $\hat{\mathbb{I}}_{2\times 2}$ and $\hat{\sigma}_3$ are the $2\times 2$ identity matrix and the third Pauli matrix, respectively.

Using equations (\ref{eq:ncsuperpotential}), (\ref{eq:Bosonic-Hamiltonian}), and (\ref{eq:fermionic_hamiltonian}), the Fermionic and Bosonic Hamiltonians can be read off as
\begin{equation}
    \begin{aligned}
        H_2&=\frac{1}{2m}\mathcal{A}*^r\mathcal{A}^\dag,\\
        H_1&=\frac{1}{2m}\mathcal{A}^\dag *^r\mathcal{A}.
    \end{aligned}
\end{equation}
The above expressions for the partner Hamiltonians then lead one to conclude that there exists a relationship between the eigenstates of $H_1$ and $H_2$. To see this relation explicitly, let us denote by $\psi_{n+1}^{(1)}$ the $(n+1)$\textsuperscript{th} excited state of the Fermionic Hamiltonian $H_1$ with energy eigenvalue denoted by $E_{n+1}^{(1)}$. One can then write
\begin{equation}
    H_1*^r\psi_{n+1}^{(1)}=E_{n+1}^{(1)}\psi_{n+1}^{(1)}.
\end{equation}
The expression above then leads one to the eigenstate of the Bosonic Hamiltonian in the following way
\begin{equation}\label{eq:partner_simultenious_eigenstate}
    \begin{aligned}
        H_2*^r\left(\mathcal{A}*^r\psi_{n+1}^{(1)}\right)&= \frac{1}{2m}\mathcal{A}*^r\mathcal{A}^\dag*^r\left(\mathcal{A}*^r\psi_{n+1}^{(1)}\right),\\
        &=\mathcal{A}*^rH_1*\psi_{n+1}^{(1)},\\
        &=\mathcal{A}*^rE_{n+1}^{(1)}\psi_{n+1}^{(1)},\\
        &=E_{n+1}^{(1)}\mathcal{A}*^r\psi_{n+1}^{(1)}.
    \end{aligned}
\end{equation}
One then concludes that $\psi_n^{(2)}=\mathcal{A}*^r\psi_{n+1}^{(1)}$ is an eigenstate of the Bosonic Hamiltonian $H_2$ with the energy eigenvalue $E_n^{(2)}=E_{n+1}^{(1)}$. Therefore, the $(n+1)$\textsuperscript{th} eigenstate of the SUSY Hamiltonian is given by
\begin{equation}\label{eq:SUSY_partner_state}
    \boldsymbol{\Psi}_{n+1}=\begin{pmatrix}
        \psi_n^{(2)}\\
        \psi_{n+1}^{(1)}
    \end{pmatrix}=\begin{pmatrix}
        \mathcal{A}*^r\psi_{n+1}^{(1)}\\
        \psi_{n+1}^{(1)}
    \end{pmatrix}.
\end{equation}

One finally arrives at the following expression of the SUSY Hamiltonian

\begin{equation}\label{eq:Schroedinger_eq}
    \begin{aligned}
        H\boldsymbol{\Psi}_{n+1}&=\begin{pmatrix}
            H_2 & 0\\
            0 & H_1
        \end{pmatrix}\begin{pmatrix}
            \psi_n^{(2)}\\
            \psi_{n+1}^{(1)}
        \end{pmatrix}=\begin{pmatrix}
            H_2*^r\psi_n^{(2)}\\
            H_1*^r\psi_{n+1}^{(1)}
        \end{pmatrix}.\\
        &=\begin{pmatrix}
            E_n^{(2)}\psi_n^{(2)}\\
            E_{n+1}^{(1)}\psi_{n+1}^{(1)}
        \end{pmatrix}.
    \end{aligned}
\end{equation}
%The expressions for the eigenvalues of the partner Hamiltonians are given by
%\begin{equation}\label{eq:eigenvalue_susy}
    %\begin{pmatrix}E_{n}^{(2)}&0\\0&E_{n}^{(1)}\end{pmatrix}=\hbar\frac{|eB|}{m}\Big(n+\frac{1}{2}\Big)\hat{\mathbb{I}}_{2\times 2} + \frac{1}{2m}\hbar eB\hat{\sigma}_3,
%\end{equation}
%for $n\geq 0$. %
%It turns out that the energy eigenvalues are all independent of $\vartheta$. 
Here, the energy eigenstates $\Psi_{n+1}=\begin{pmatrix}
        \psi_n^{(2)}\\
        \psi_{n+1}^{(1)}
    \end{pmatrix}\in \mathcal{F}=((C^{\infty}(\mathbb{R}^{2})\cap L^{2}(\mathbb{R}^{2},\hbox{dx}\;\hbox{dy}))[[\vartheta]],*^r)\oplus((C^{\infty}(\mathbb{R}^{2})\cap L^{2}(\mathbb{R}^{2},\hbox{dx}\;\hbox{dy}))[[\vartheta]],*^r)$, for $n\geq0$. One immediately observes using (\ref{eq:eigenvalue_susy}) and (\ref{eq:partner_simultenious_eigenstate}) that the component wavefunctions $\psi_{n}^{(2)}=\mathcal{A}*^r\psi_{n+1}^{(1)}$ and $\psi_{n+1}^{(1)}$ of the SUSY eigenstate $\Psi_{n+1}=\begin{pmatrix}
        \mathcal{A}*^r\psi_{n+1}^{(1)}\\
        \psi_{n+1}^{(1)}
    \end{pmatrix}$, are indeed eigenstates of $H_2$ and $H_1$, respectively, with the same eigenvalue 
    
    \begin{equation}\label{eq:Eigenvalues_Hamiltonian}
    E_{n+1}^{(1)}=E_{n}^{(2)}=\hbar\frac{eB}{m}(n+1)=:E_{n+1}^{\hbox{\tiny{susy}}},\quad \hbox{for}\;\; n\geq 0.
    \end{equation}
    It turns out that the energy eigenvalues are all independent of the noncommutativity parameter $\vartheta$ and the gauge parameter $r$.
    Hence, the eigenvalue equation (\ref{eq:Schroedinger_eq}) for the SUSY Hamiltonian $H$ reduces to 
    \begin{equation}
H\Psi_{n+1}=E_{n+1}^{\hbox{\tiny susy}}\Psi_{n+1},
    \end{equation}
for $n\geq 0$, where $E_{n+1}^{\hbox{\tiny {susy}}}$ is given by (\ref{eq:Eigenvalues_Hamiltonian}).

From (\ref{eq:eigenvalue_susy}), one can see that the ground state energy $E^{(1)}_{0}$ of the Fermionic Hamiltonian $H_1$ is indeed zero. There is no nontrivial eigenstate of the Bosonic Hamiltonian $H_2$ associated with zero energy. In other words, the ground state $\Psi_{0}$ of the SUSY Hamiltonian has the form:
\begin{equation}\label{eq:ground-state-SUSY-Hamiltonian}
    \Psi_{0}=\begin{pmatrix}
    0\\
    \psi_0^{(1)}
    \end{pmatrix}.
\end{equation}
Hence, one concludes that the supersymmetry remains unbroken in all orders of $\vartheta$ for the model we are considering.

From equation (\ref{eq:eigenvalue_susy}), it is evident that the ground state energy of the Fermionic Hamiltonian is zero, which amounts to
\begin{equation}
    H_1*^r\psi_0^{(1)}=\frac{1}{2m}\mathcal{A}^\dag*^r\mathcal{A}*^r\psi_0^{(1)}=0.
\end{equation}
One then concludes from the above equation that the ground state $\psi^{(1)}_{0}$, of the Fermionic Hamiltonian $H_1$, is $*^{r}$-annihilated by the superpotential $\mathcal{A}$, i.e.,
\begin{equation}\label{eq:ground_state_eq}
    \mathcal{A}*^r\psi_0^{(1)}=0.
\end{equation}
One can use standard techniques of seperation of variables for solving partial differential equations to achieve an explicit expression of $\psi_0^{(1)}$. The detailed calculation is given in the Appendix (\ref{appendix:ground_state_of_fermionic_hamiltonian}). In what follows, we write the gauge parameter $r$-dependent final expression of the ground state
\begin{equation}\label{eq:r-dependent-wave-function-main}
    \begin{aligned}
        \psi_{0,r}^{(1)}(x,y)&= \exp\Bigg[\frac{1}{\hbar\left\{1+\frac{2r(1-r)e\vartheta B}{\hbar +\sqrt{\hbar^2-4r(r-1)e\hbar\vartheta B}}\right\}}\Big\{mx+imy-\frac{re\hbar Bx^2}{\hbar +\sqrt{\hbar^2-4r(r-1)e\hbar\vartheta B}}\\
        &\qquad\qquad-\frac{(1-r)e\hbar By^2}{\hbar +\sqrt{\hbar^2-4r(r-1)e\hbar\vartheta B}}\Big\}+k\Bigg],
    \end{aligned}
\end{equation}
where $k$ is some constatnt determined by the boundary conditions imposed on the wavefunction. Plugging in $r=0$, one obtains the Fermionic ground state in the Landau gauge given by
\begin{equation}\label{eq:Landau-ground-state}
    \psi_{0,\hbox{\tiny{Lan}}}^{(1)}(x,y)=\exp\left[\frac{1}{\hbar}\left(mx+imy-\frac{eBy^2}{2}\right)\right].
\end{equation}
Similarly, by setting $r=\frac{1}{2}$, one gets the Fermionic ground state in the symmetric gauge given by
\begin{equation}\label{eq:symmetric-ground-state}
    \begin{aligned}
        \psi&_{0,\hbox{\tiny{Sym}}}^{(1)}(x,y)\\
        &=\exp\left[\frac{1}{\hbar\left\{1+\frac{e\hbar \vartheta B}{2\hbar+2\sqrt{\hbar^2+e\hbar\vartheta B}})\right\}}\left\{mx+imy-\frac{e\hbar B}{2\hbar+2\sqrt{\hbar^2+e\hbar\vartheta B}}(x^2+y^2)\right\}+c\right].
    \end{aligned}
\end{equation}

The formalism that we have developed so far in this section is a gauge invariant formalism in the sense that one can change the value of the gauge parameter $r$ to consider the SUSY system in different gauges but there is no effect of the gauge parameter $r$ in the energy eigenvalues of the SUSY Hamiltonian given by (\ref{eq:Eigenvalues_Hamiltonian}). Now we proceed to compute the n\textsuperscript{th} excited state of the Fermionic Hamiltonian $H_1$ in the familiar Landau and Symmetric gauges and verify that the respective energy eigenvalues are indeed the same. With the Fermionic Hamiltonians in both the gauges at our disposal, we then construct the n\textsuperscript{th} eigenstate of the SUSY Hamiltonian in both Landau and Symmetric gauges.

Let us address the Landau gauge first. Plugging in $r=0$ in the second equation of (\ref{eq:partner_hamiltonian}), one obtains the expression for the algebra element of the Fermionic Hamiltonian $H^{\hbox{\tiny{Lan}}}_{1}$ in the Landau gauge. Going back to the operatorial presentation, the corresponding Fermionic Hamiltonian operator $\hat{H}^{\hbox{\tiny{Lan}}}_{1}$ in the Landau gauge can be expressed as
\begin{equation}\label{eq:Fermionic_Hamiltonian_in_Landau_gauge}
    \begin{aligned}
        \hat{H}^{\hbox{\tiny{Lan}}}_{1} &=\frac{1}{2m}\left[\hat{p}_x^2+(\hat{p}_y-eB\hat{x})(\hat{p}_y+eB\hat{x})-e\hbar B\hat{\mathbb{I}}\right],\\
        &=\frac{1}{2m}\left[\hat{p}_x^2+(\hat{p}_y-eB\hat{x})^2+e\hbar B\hat{\mathbb{I}}\right].
    \end{aligned}
\end{equation}
The Hamiltonian operator given by the above equation consists of two parts as discussed before. The first part given by
\begin{equation}\label{eq:l_hamiltonian}
    \frac{1}{2m}\left[\hat{p}_x^2+(\hat{p}_y-eB\hat{x})^2\right],
\end{equation}
represents the Landau Hamiltonian operator in the Landau gauge. The second part $\frac{1}{2m}e\hbar B\hat{\mathbb{I}}$ is an extra contribution. These two parts of the Hamiltonian operator admit simultaneous eigenstates which is precisely the eigenstate of the Landau Hamiltonian operator given by (\ref{eq:l_hamiltonian}). To find the explicit expression of this eigenstate, one should first notice that the operator $\hat{p}_y$ commutes with the Landau Hamiltonian operator given in (\ref{eq:l_hamiltonian}). Therefore, one can replace $\hat{p}_y$ with its eigenvalue $k_y$ times the identity operator $\hat{\mathbb{I}}$ on $L^{2}(\mathbb{R}^{2},\hbox{dx}\,\hbox{dy})$ in (\ref{eq:l_hamiltonian}). The equation (\ref{eq:l_hamiltonian}) then reduces to
\begin{equation}
    \frac{1}{2m}\left[\hat{p}_x^2+(k_y\hat{\mathbb{I}}-eB\hat{x})^2\right].
\end{equation}
One can further simplify the above expression as
\begin{equation}
    \frac{\hat{p}_x^2}{2m}+\frac{1}{2}m\frac{e^2B^2}{m^2}\left(\hat{x}-\frac{k_y}{eB}\hat{\mathbb{I}}\right)^2.
\end{equation}
The expression above is that of the Hamiltonian operator of a 1-dimensional quantum harmonic oscillator with the potential shifted in the coordinate space by $\frac{k_y}{eB}$ and the angular frequency $\frac{eB}{m}$. The energy eigenvalues and the n\textsuperscript{th} excited eigenstate of the Landau Hamiltonian operator (which is also an eigenstate of the Fermionic Hamiltonian since the Landau Hamiltonian operator commutes with the Fermionic Hamiltonian operator) (see \ref{eq:l_hamiltonian} and \ref{eq:Fermionic_Hamiltonian_in_Landau_gauge}) are given by
\begin{equation}\label{eq:fermionic_excited_state_land}
    \begin{aligned}
    E_n^{\hbox{\tiny Lan}}&=\hbar\frac{eB}{m}\left(n+\frac{1}{2}\right)\;\;\hbox{and},\\
    \psi_{n,\hbox{\tiny Lan}}^{(1)}(x,y)&=N e^{ik_yy}\exp\left\{-\frac{eB}{2\hbar}\left(x-\frac{k_y}{eB}\right)^2\right\}\mathcal{H}_n\left(\sqrt{\frac{eB}{\hbar}}\left(x-\frac{k_y}{eB}\right)\right),
    \end{aligned}
\end{equation}
respectively, where $N$ is a suitably chosen normalization constant and $\mathcal{H}_n$ is the $n$-th Hermite polynomial in the shifted $x$-coordinate. Hence, the energy eigenvalues of the Fermionic Hamiltonian operator $\hat{H}^{\hbox{\tiny{Lan}}}_1$, given by (\ref{eq:Fermionic_Hamiltonian_in_Landau_gauge}), in the Landau gauge are given by
\begin{equation}\label{eq:energy_landau_detail}
    \begin{aligned}
        E_{n,\hbox{\tiny Lan}}^{(1)}&=\left(n+\frac{1}{2}\right)\hbar\frac{eB}{m}-\frac{e\hbar B}{2m},\\
        &=\hbar\frac{eB}{m}n,
    \end{aligned}
\end{equation}
while, using (\ref{eq:SUSY_partner_state}), the $n$\textsuperscript{th} eigenstate of the SUSY Hamiltonian can be read off as
\begin{equation}\label{eq:SUSY-eigenstate-Landau-gauge}
    \boldsymbol{\Psi}_n^{\hbox{\tiny Lan}}(x,y)=\begin{pmatrix}
        \hat{\mathcal{A}}^0\psi_{n,\hbox{\tiny Lan}}^{(1)}(x,y)\\
        \psi_{n,\hbox{\tiny Lan}}^{(1)}(x,y)
    \end{pmatrix},
\end{equation}
with $\hat{\mathcal{A}}^0$ being the superpotential operator associated with the superpotential $\mathcal{A}\in (C^{\infty}(\mathbb{R}^{2})[[\vartheta]],*^{r})$ (see (\ref{eq:ncsuperpotential})) expressed in the Landau gauge.

The above calculation is done purely in the Landau gauge, which, in this article, is achieved by setting the gauge parameter $r$ to zero. In what follows next, let us show that the same energy eigenvalues can be attained if one considers the Fermionic Hamiltonian in the symmetric gauge by setting the value of the gauge parameter $r$ to $\frac{1}{2}$.

To express the Fermionic Hamiltonian (\ref{eq:partner_hamiltonian}) in the symmetric gauge, let us first evaluate
\begin{equation}
    \begin{aligned}
        (p_x&-e\boldsymbol{A}_x^{\hbox{\tiny nc}})*^{\frac{1}{2}}\psi_{n,\hbox{\tiny Sym}}^{(1)}=\left(\hat{p}_x+\frac{e\hbar B}{\hbar+\sqrt{\hbar^2+e\hbar \vartheta B}}\hat{Y}^{\frac{1}{2}}\right)\psi_{n,\hbox{\tiny Sym}}^{(1)},\\
        &=\left[\hat{p}_x+\frac{e\hbar B}{\hbar+\sqrt{\hbar^2+e\hbar \vartheta B}}\left(\hat{y}+\frac{\vartheta}{2\hbar}\hat{p}_x\right)\right]\psi_{n,\hbox{\tiny Sym}}^{(1)},\\
        &=\left[1+\frac{e\vartheta B}{2(\hbar+\sqrt{\hbar^2+e\hbar \vartheta B})}\right]\left[\hat{p}_x+\frac{e}{1+\frac{e\vartheta B}{2(\hbar+\sqrt{\hbar^2+e\hbar \vartheta B})}}\frac{\hbar B}{\hbar+\sqrt{\hbar^2+e\hbar \vartheta B}}\hat{y}\right]\psi_{n,\hbox{\tiny Sym}}^{(1)},
    \end{aligned}
\end{equation}
by setting $r=\frac{1}{2}$ for the symmetric gauge.

Similarly, one proceeds to show that the following holds
\begin{equation}
    \begin{aligned}
        (\hat{p}_y&-e\boldsymbol{A}_y^{\hbox{\tiny{nc}}})*^{\frac{1}{2}}\psi_{n,\hbox{\tiny Sym}}^{(1)}\\
        &=\left[1+\frac{e\vartheta B}{2(\hbar+\sqrt{\hbar^2+e\hbar \vartheta B})}\right]\left[\hat{p}_y-\frac{e}{1+\frac{e\vartheta B}{2(\hbar+\sqrt{\hbar^2+e\hbar \vartheta B})}}\frac{\hbar B}{\hbar+\sqrt{\hbar^2+e\hbar \vartheta B}}\hat{x}\right]\psi_{n,\hbox{\tiny Sym}}^{(1)}.
    \end{aligned}
\end{equation}
Let us denote
\begin{equation}\label{eq:deformed-quantities}
    \begin{aligned}
        \bar{\Lambda}\left(\frac{1}{2}\right)&=1+\frac{e\vartheta B}{2(\hbar+\sqrt{\hbar^2+e\hbar \vartheta B})},\\
        e_*&=\frac{e}{\bar{\Lambda}\left(\frac{1}{2}\right)},\\
        m_*&=\frac{m}{\bar{\Lambda}^2\left(\frac{1}{2}\right)},\\
        \bar{B}&=\frac{\hbar B}{\hbar+\sqrt{\hbar^2+e\hbar\vartheta B}}.
    \end{aligned}
\end{equation}
The notation $\bar{\Lambda}$ is borrowed from \cite{Chowdhury1} where $\bar{\Lambda}(r)$ as a function of the gauge parameter $r$ is provided in equation (3.14) on p. 8 of \cite{Chowdhury1}. Therefore, using the above notations, one can rewrite the Fermionic Hamiltonian operator in the symmetric gauge as
\begin{equation}
    \hat{H}_1^{\hbox{\tiny sym}}=\frac{1}{2m_*}(\hat{p}_x^2+\hat{p}_y^2)+\frac{1}{2}m_*\left(\frac{e_*\bar{B}}{m_*}\right)^2(\hat{x}^2+\hat{y}^2)-\left(\frac{e_*\bar{B}}{m_*}\right)(\hat{x}\hat{p}_y-\hat{y}\hat{p}_x)-\frac{1}{2}\hbar\frac{eB}{m}\hat{\mathbb{I}}.
\end{equation}
One also obtains
\begin{equation}
    \frac{e_*\bar{B}}{m_*}=\frac{e\bar{\Lambda}(\frac{1}{2})\bar{B}}{m}=\frac{eB}{2m}.
\end{equation}
Let us define
\begin{equation}
    \omega_c=\frac{eB}{m}.
\end{equation}
One finally arrives at the following expression of the Fermionic Hamiltonian operator in the symmetric gauge
\begin{equation}\label{eq:Fermionic-Hamiltonian-symmetric-gauge}
    \hat{H}_1^{\hbox{\tiny sym}}=\frac{1}{2m_*}(\hat{p}_x^2+\hat{p}_y^2)+\frac{1}{2}m_*\left(\frac{\omega_c}{2}\right)^2(\hat{x}^2+\hat{y}^2)-\left(\frac{\omega_{c}}{2}\right)(\hat{x}\hat{p}_y-\hat{y}\hat{p}_x)-\frac{1}{2}\hbar\omega_c\hat{\mathbb{I}}.
\end{equation}
The Fermionic Hamiltonian operator $\hat{H}_1^{\hbox{\tiny sym}}$, in the symmetric gauge, consists of two parts. The first one, being
\begin{equation}\label{eq:Landau-Hamiltonian-symmetric-gauge}
    \frac{1}{2m_*}(\hat{p}_x^2+\hat{p}_y^2)+\frac{1}{2}m_*\left(\frac{\omega_c}{2}\right)^2(\hat{x}^2+\hat{y}^2)-\left(\frac{\omega_c}{2}\right)\hat{L}_z,
\end{equation}
with $\hat{L}_z=\hat{x}\hat{p}_y-\hat{y}\hat{p}_x$, resembles exactly the Landau Hamiltonian operator in the symmetric gauge. The rest is an extra contribution arising from the Pauli equation.
These two pieces of the Hamiltonian operator commute with each other resulting in the fact that the Fermionic Hamiltonian operator $\hat{H}_1^{\hbox{\tiny sym}}$ (see \ref{eq:Fermionic-Hamiltonian-symmetric-gauge}) in the symmetric gauge and the Landau Hamiltonian operator (see \ref{eq:Landau-Hamiltonian-symmetric-gauge}) expressed in the symmetric gauge share simultaneous eigenstates. Therefore, it suffices to evaluate the eigenstates of the Landau Hamiltonian operator in the symmetric gauge given by (\ref{eq:Landau-Hamiltonian-symmetric-gauge}). The most convenient way of solving the eigenvalue equation for the Landau Hamiltonian operator in the symmetric gauge is to use the cylindrical coordinate system. The excited eigenstates of the pertaining Landau Hamiltonian operator in the symmetric gauge can be read off as
\begin{equation}\label{eq:fermionic_excited_state_symmetric}
    \begin{aligned}
        \psi_{n,\hbox{\tiny Sym}}^{(1)}(\rho,\phi)= &N_{n_\rho,m_l}\left(\sqrt{\frac{\bar{\Lambda}(\frac{1}{2})eB}{2\hbar}}\rho\right)^{|m_l|}\exp(-\frac{\rho^2\bar{\Lambda}(\frac{1}{2})eB}{4\hbar})\\
        &\qquad L_{n_\rho}^{|m_l|}\left(\frac{\rho^2\bar{\Lambda}(\frac{1}{2})eB}{2\hbar}\right)\exp(im_l\phi),
    \end{aligned}
\end{equation}
where $N_{n_\rho,m_l}$ is a normalization factor, $L_{n_\rho}^{|m_l|}$ is the associated Laguerre polynomial and $\rho^2=x^2+y^2$. The relationship of the quantum number $n$, attached to the Fermionic excited eigenstate $\psi_{n,\hbox{\tiny Sym}}^{(1)}$ in the symmetric gauge, with $n_\rho$ and $m_l$ appearing on the right side of (\ref{eq:fermionic_excited_state_symmetric}) will be explored shortly. One, therefore, writes the  $n$\textsuperscript{th} eigenstate (see (\ref{eq:SUSY_partner_state})) of the SUSY Hamiltonian in the symmetric gauge as 
\begin{equation}\label{eq:SUSY-eigenstate-symmetric-gauge}
    \boldsymbol{\Psi}_n^{\hbox{\tiny Sym}}(\rho,\phi)=\begin{pmatrix}
        \hat{\mathcal{A}}^{\frac{1}{2}}\psi_{n,\hbox{\tiny Sym}}^{(1)}(\rho,\phi)\\
        \psi_{n,\hbox{\tiny Sym}}^{(1)}(\rho,\phi)  
        \end{pmatrix},
\end{equation}
where $\hat{\mathcal{A}}^{\frac{1}{2}}$ is the superpotential operator corresponding to the superpotential $\mathcal{A}\in (C^{\infty}(\mathbb{R}^{2})[[\vartheta]],*^{r})$ (see (\ref{eq:ncsuperpotential})) expressed in the symmetric gauge. 

The energy eigenvalues of the Landau Hamiltonian operator in the symmetric gauge are given by
\begin{equation}\label{eq:Landau_Hamiltonian_Degeneracy}
    E_{n_\rho,m_l}^{\hbox{\tiny Sym}}=\hbar\frac{eB}{m}\left(n+\frac{1}{2}\right),
\end{equation}
where $n$ is given by
\begin{align}
    \label{eq:Fermionic_Hamiltonian_Degeneracy_nonnegative}
    n=n_{\rho}+\frac{|m_{l}|-m_{l}}{2}
    =\begin{cases}
        n_\rho, \quad \qquad \quad\text{ if } m_l \text{ is a nonnegative integer,}\\
        n_\rho+|m_l|, \quad \text{ if } m_l \text{ is a negative integer.}
    \end{cases}
\end{align}
The details of these calculations can be found in \cite{Ciftja}. Finally, the energy eigenvalues of the Fermionic Hamiltonian operator $\hat{H}_1^{\hbox{\tiny sym}}$ in the symmetric gauge can be read off as
\begin{equation}\label{eq:energy_symmetric_detail}
    E_{n,\hbox{\tiny Sym}}^{(1)}=\hbar\frac{eB}{m}n.
\end{equation}
which exactly agrees with the expression for the energy eigenvalues given by (\ref{eq:energy_landau_detail}). It is important to note here, using(\ref{eq:Fermionic_Hamiltonian_Degeneracy_nonnegative}), that the energy eigenvalues given by (\ref{eq:Landau_Hamiltonian_Degeneracy}) are degenerate in the non-negative integer values of $m_l$, contributing to the degeneracy in the eigenvalues of the Fermionic Hamiltonian. To be more specific, any Fermionic excited eigenstate (see \ref{eq:fermionic_excited_state_symmetric}), in the symmetric gauge corresponding to a given integer value of the radial quantum number $n_{\rho}$, will yield the same energy eigenvalue (which is equal to $\frac{\hbar eBn_{\rho}}{m}$) for whatever non-negative integer value we choose for the angular quantum number $m_{l}$.

%--------------------- Comparison of Results and Discussions --------------%

\section{Comparison with Other Results and Discussion}\label{sec:comparison}

In this article, we have adopted a mathematically and physically consistent approach to deal with the $\mathcal{N}=2$ supersymmetric Landau problem in a 2-dimensional plane using the deformation quantization technique developed in \cite{Chowdhury1}. It turns out that the energy eigenvalues of the supersymmetric Hamiltonian is independent of the noncommutativity parameter $\vartheta$ (see (\ref{eq:eigenvalue_susy})). It is a physically consistent formalism as the energy eigenvalues of the supersymmetric Hamiltonian one obtains are all independent of the gauge parameter $r$. In particular, all the energy eigenvalues of the bosonic and fermionic partner Hamiltonians are the same in both Landau and symmetric gauges. The energy eigenstates, on the other hand, for the partner Hamiltonians are indeed gauge dependent (gauge parameter $r$ dependent) expressions. We have explicitly computed an $r$-dependent continuously degenerate ground state of the fermionic Hamiltonian in (\ref{eq:r-dependent-wavefunc}). Plugging in $r=1$ and $r=\frac{1}{2}$ yield the ground state of the fermionic Hamiltonian in the Landau and symmetric gauges, respectively (see \ref{eq:Landau-ground-state},\ref{eq:symmetric-ground-state}). 

To compare our results with the existing literature on supersymmetric Landau problem in a noncommutative plane, first note that in \cite{Harikumaretal}, the authors claim that the Seiberg-Witten map in the first order yields the same energy spectra with the noncommutativity correction incorporated in it both in Landau and symmetric gauges. In addition to this, they prove that supersymmetry remains unbroken there in the first order of $\theta$. There they obtain the same energy eigenvalue in the first order of $\theta$ for both the symmetric and the Landau gauges (see equation (33) and (35) on p. 160) because of the expression of the noncommutative gauge fields in symmetric gauge given by equation (29) on p.159 in their article. It is important to note that equation (29) there can not be derived from the first order Seiberg-Witten map given by equation (9) appearing on p. 159. What actually follows from equation (9) on p. 159 there for the symmetric gauge in the first order of $\theta$ are given in the appendix (\ref{eq:A_hat_x},\ref{eq:A_hat_y}). These expressions lead one to the supersymmetric Hamiltonian (\ref{eq:SUSY-Hamiltonian_Symmetric}) in the appendix as opposed to equation (34) of \cite{Harikumaretal} on p. 160. The resulting SUSY Hamiltonian in the symmetric gauge then yields the eigenvalues (\ref{eq:eigenvalues-symmetric-Harikumaretal}) in contrast to what were obtained in equation (35) of p. 160 in \cite{Harikumaretal}. Equation (\ref{eq:ground-state-energy-Landau-gauge}) obtained from (\ref{eq:eigenvalues-symmetric-Harikumaretal}) clearly indicates that the ground state energy in the symmetric gauge becomes non-zero violating the conclusions claimed in \cite{Harikumaretal}. The resulting anomaly associated with the nonzero ground state energy of the SUSY Hamiltonian in the symmetric gauge is due to the fact that the noncommutative gauge fields were obtained from the commutative ones by limiting only to the first order (in $\theta$) Seiberg-Witten map (see equation (9) on p. 159). In contrast to (\cite{Harikumaretal}), we resort to the exact expression of the noncommutative gauge fields in terms of the commutative gauge fields given by (\ref{eq:nc-gaugefields-exact-expression}) to obtain a gauge invariant energy spectra (\ref{eq:eigenvalue_susy}) of the underlying SUSY Hamiltonian (\ref{eq:SUSY-Hamiltonian-final-form}). It is worth remarking in this context that there is indeed a formal power series expansion in $\vartheta$ of the exact expression of the gauge fields (\ref{eq:nc-gaugefields-exact-expression}) given by equations (4.10) and (4.14) on p. 13 of \cite{Chowdhury1}. And these expressions align with the ones given by the familiar Seiberg-Witten map when one sets the gauge parameter $r=\frac{1}{2}$ (see equation (4.20) of \cite{Chowdhury1} on p. 14). From this viewpoint, if one takes the Seiberg-Witten map into account to transform the commutative gauge fields into the noncommutative ones, one has to take the expressions containing all orders of $\vartheta$ in order to obtain a gauge invariant formulation of supersymmetric quantum mechanics in a noncommutative plane.

The authors in \cite{Halderetal}, on the other hand, claimed to have used the Seiberg-Witten map in all orders of the noncommutativity parameter $\theta$ to address the solutions of the Pauli equation associated with a nonrelativistic spin $\frac{1}{2}$ charged particle placed in a noncommutative plane under the influence of a vertical uniform magnetic field. There they claim to have obtained the energy eigenvalues (equation (46) on p. 11 of \cite{Halderetal}) of the Pauli Hamiltonian in all order of $\theta$. The problem arises when one takes $n=0$ in that expression. The ground state energy is not zero there either in the symmetric gauge leading to the fact that the underlying supersymmetry is broken in all order of $\theta$. This is in sharp contrast to what we have found in the present paper (see the discussions around equation (\ref{eq:ground-state-SUSY-Hamiltonian})) that supersymmetry remains unbroken in all order of the noncommutativity parameter $\vartheta$ (note that we have used $\vartheta$ to denote the spatial noncommutativity parameter while the authors in \cite{Ashokdasetal, Harikumaretal, Halderetal} used $\theta$ to denote the same). 

The authors in \cite{Ashokdasetal} used the Moyal product to obtain a deformed SUSY algebra (equation (32) on p. 410 of \cite{Ashokdasetal}) to formulate quantum mechanics in noncommutative spaces. They started the treatment with the case of a simple noncommutative manifold in $2$ spatial dimensions. There they used Bopp shift (a consequence of the Moyal product) to transform the commutative spatial coordinates into the noncommutative ones. And to elucidate the general construction further, they took the example of the SUSY Landau problem associated with a noncommutative $2$-plane. Then they proceed to compute the eigenvalues of the SUSY Hamiltonian in the symmetric gauge (given by equation (40) on p.411 of \cite{Ashokdasetal}). The energy eigenvalues have noncommutative correction in the expression they obtained (see equations (46) and (51) of \cite{Ashokdasetal}). The verification of the calculations are all provided in detail in appendix (\ref{susec:appendix-ashokdasetal}) for the sake of completeness. The gauge prescription the authors adopted in this example of SUSY Landau problem for noncommutative $2$-plane is termed as naive minimal prescription in \cite{Chowdhury2}. It has been proved there that such gauge prescription yields gauge dependent energy spectra for noncommutative two dimensional anisotropic harmonic oscillator in a constatnt magnetic field (see sec 3.3 on p. 11 of \cite{Chowdhury2}). We repeated the calculation undertaken in \cite{Ashokdasetal} for the Landau gauge (see (\ref{eq:landau-gauge-ashokdasetal})) following naive minimal prescription in appendix (\ref{susec:appendix-ashokdasetal}). The energy eigenvalues (see (\ref{eq:eigenvalues-landau-gauge-SUSY-Hamiltonian-Naive-Minimal-prescription})) then obtained do not have the noncommutativity parameter 
$\theta$ in their expression as opposed to the symmetric gauge expression (see (\ref{eq:symmetric-gauge-eigenvalue-naive-minimal-prescription})). In other words, the energy eigenvalues of the SUSY Hamiltonian associated with the SUSY Landau problem in a noncommutative $2$-plane that the authors in \cite{Ashokdasetal} obtained are gauge dependent. It is worth remarking at this stage that despite the gauge dependency of the energy eigenvalues of the SUSY Hamiltonian, the underlying ground state energy is found to be zero both in the symmetric and the Landau gauges there which points up the fact that supersymmetry remains unbroken in this setup.

\section{Conclusion and Outlook}\label{sec:conclusion}
In this article, we have studied the quantum mechanical system of a spin $1/2$ charged particle situated in a noncommutative plane under the influence of a uniform vertical magnetic field using the technique of deformation quantization. The noncommutative SUSY Landau model that we are considering here is found to respect the deformed $\mathcal{N}=2$ supersymmetry algebra (see (\ref{eq:deformed-SUSY-algebra-N=2}) and (\ref{eq:paraity-commutation-relation})). The eigenvalues of the SUSY Hamiltonian $H$ turn out to be all not only independent of the gauge parameter $r$ but also independent of the noncommutativity parameter $\vartheta$ (see (\ref{eq:Eigenvalues_Hamiltonian})). The gauge parameter $r$-dependent nontrivial ground state (see \ref{eq:r-dependent-wave-function-main}) of the Fermionic Hamiltonian corresponds to zero energy and hence the SUSY remains unbroken in all order of $\vartheta$. This result is in sharp contrast to the findings of \cite{Harikumaretal,Halderetal} as discussed at length in section \ref{sec:comparison}. Finally, the excited states of the partner Hamiltonians were computed both in Landau and symmetric gauges. The $n$\textsuperscript{th} excited state (see \ref{eq:fermionic_excited_state_land}) of the Fermionic Hamiltonian $H_1$ in the Landau gauge has no $\vartheta$ contribution while the one (see (\ref{eq:fermionic_excited_state_symmetric})) in the symmetric gauge has $\vartheta$ dependence through the quantity $\bar{\Lambda}\left(\frac{1}{2}\right)$ defined in (\ref{eq:deformed-quantities}).

Witten introduced a topological characteristic associated with the ground state of the SUSY Hamiltonian known as the {\em Witten index} \cite{Witten_constraints}. The Witten index is defined by
\begin{equation}\label{eq:Witten-index-def-groundstate}
    \Delta = n_2-n_1,
\end{equation}
where $n_2$ and $n_1$ are the numbers of zero energy Bosonic states and the number of zero energy Fermionic states, respectively. These two positive integers $n_1$ and $n_2$ can be interpreted as the dimension of the kernel of the superpotential $\mathcal{A}$ and its adjoint $\mathcal{A}^\dag$, respectively. Therefore, the Witten index $\Delta$ given in (\ref{eq:Witten-index-def-groundstate}) can be recast into the following expression
\begin{equation}\label{eq:Witten-index-elliptic-operator}
    \Delta = \dim\ker \mathcal{A}^\dag-\dim \ker \mathcal{A}.
\end{equation}
To determine  $\dim\ker \mathcal{A}$ and $\dim\ker \mathcal{A}^\dag$, one needs to focus on the solution space of the equations $\mathcal{A}*^r\phi=0$ and $\mathcal{A}^\dag *^r\phi=0$, where by $0$, we mean the constant $0$-function in $C^{\infty}(\mathbb{R}^{2})$ that maps all of $\mathbb{R}^{2}$ to the constant real number $0$. By referring back to (\ref{eq:ground_state_eq}) and the appendix (\ref{appendix:ground_state_of_fermionic_hamiltonian}), one can conclude that the only nontrivial solutions of the homogeneous equation $\mathcal{A}*^r\phi=0$ are the complex multiples of the ground state wave function of the Fermionic Hamiltonian $H_1$ given by (\ref{eq:r-dependent-wave-function-main}) which refers to the fact that $\dim\ker\mathcal{A}=1$. Besides, the equation (\ref{eq:SUSY_partner_state}) indicates that there is no nontrivial ground state of the Bosonic Hamiltonian $H_2=\frac{1}{2m}\mathcal{A}*^{r}\mathcal{A}^\dag$. Now, $\ker\mathcal{A}^{\dag}\subset\ker H_{2}$ together with $\ker H_{2}=\{0\}$ implies that $\ker\mathcal{A}^{\dag}=\{0\}$. One then immediately concludes that $\dim\ker\mathcal{A}^\dag=0$. Hence, the Witten index $\Delta = -1\neq 0$. The calculated Witten index coincides with the result for the SUSY Landau problem in the commutative setup ($\vartheta=0$). This result also points up the fact that there is no broken supersymmetry for the model of the SUSY Landau problem in the noncommutative plane we are considering (see page 257, \cite{Witten_constraints}). 

A couple of directions can be pursued following what have been done in this article from the deformation quantization perspective. We plan to study relativistic SUSY Landau problem in a noncommutative plane by studying the relevant Dirac Hamiltonian (consult p. 9-12 of \cite{Junker} for the one in the commutative setup). For example, the Hamiltonian modelling a massless Dirac Fermion in a uniform magnetic field was considered in \cite{Hughesetal}. We expect to obtain results conforming with the ones in \cite{Hughesetal} as the noncommutativity parameter $\vartheta$ approaches zero. Another direction involves considering a spin $\frac{1}{2}$ charged particle placed on a fuzzy $2$-sphere with a Dirac monopole situated at the center of it. In this proposed model, the analogue of the noncommutative 2-plane considered in this article is the fuzzy 2-sphere. Since the phase space for a particle moving in a 2-dimensional plane is $\mathbb{R}^{4}$ and $\mathbb{R}^{4}$ possesses translational symmetry, it was natural to centrally extend the abelian group of translations in $\mathbb{R}^{4}$ by $\mathbb{R}^{3}$ (since we demanded position-position and momentum-momentum noncommutativity in addition to the quantum mechanical canonical commutation relations to hold) to construct the desired projective representations of it. The analogous Lie group to consider for the proposed model will be the semidirect product group $\mathbb{R}^{3}\rtimes \hbox{SO}(3)$ that acts transitively on the phase space $\hbox{T}^{*}\hbox{S}^{2}$, the cotangent bundle on the 2-sphere as discussed in \cite{Silvaetal}. The idea will then be to carry out the construction proposed in \cite{Chowdhury1} and in the present article by carefully studying the unitary irreducible representations of the Euclidean group $\mathbb{R}^{3}\rtimes \hbox{SO}(3)$. We then expect to reproduce the results given in \cite{D'Hokeretal} as $\vartheta\rightarrow 0$.

\subsubsection*{Acknowledgement}
This work was supported by TWAS under the research grant no. $22$-$213$ RG/MATHS/AS\textunderscore I-FR $3240324946$. The authors would like to thank Hishamuddin Zainuddin for his insightful discussions on the prospects of the project. Special thanks go out to Tibra Ali and Tushar Mitra for the valuable instructions they provided to calculate the eigenvalues of the relevant Hamiltonians. The authors also deeply acknowledge fruitful discussions with Onirban Islam and Mohammed Shadman Salam that helped them understand the unbroken supersymmetry and Witten operator.

\section{Appendix}

In the first two appendices, we gather all the detailed calculations together associated with the claims we made in this article with reference to the articles \cite{Harikumaretal, Ashokdasetal}. It is important to note that many of the notations in the articles mentioned above are confusing. Despite that we stick to the original notations in order to facilitate the comparisons mentioned in section \ref{sec:comparison}. In the last appendix, the details of the calculation for the ground state of the Fermionic Hamiltonian $H_1$ are incorporated.

\subsection{Detailed Calculation associated with the article \texorpdfstring{\cite{Harikumaretal}}{Lg}}\label{app:Harikumar}

To begin with, let us consider the symmetric gauge
\begin{equation}
    A_x=-\frac{1}{2}By,\qquad A_y=\frac{1}{2}Bx.
\end{equation}

According to the paper, one uses equation (9) to calculate the noncommutative potential $\hat{A}_x$ and $\hat{A}_y$. Equation (9) is given by
\begin{equation}\label{eq:sw}
    \hat{A}_i=A_i-\frac{1}{2}\epsilon^{kl}\theta A_k(\partial_lA_i+F_{li}).
\end{equation}
For the simplification of the calculation, let us consider $x=1$ and $y=2$. Therefore, one can write
\begin{equation}\label{eq:A_hat_x}
    \begin{aligned}
        \hat{A}_{1,\hbox{\tiny sym}} &= A_1-\frac{1}{2}\epsilon^{12}\theta A_1(\partial_2A_1+F_{21})-\frac{1}{2}\epsilon^{21}\theta A_2(\partial_1A_1+F_{11}),\\
        &= -\frac{1}{2}By-\frac{1}{2}\theta \left(-\frac{1}{2}By\right)\left(-\frac{1}{2}B-B)\right),\\
        &=-\frac{1}{2}By-\frac{3}{8}\theta B^2y,\\
        &=-\frac{1}{2}B\left(1+\frac{3}{4}B\theta\right)y.
    \end{aligned}
\end{equation}
Similarly,
\begin{equation}\label{eq:A_hat_y}
    \begin{aligned}
        \hat{A}_{2,\hbox{\tiny sym}}&=A_2-\frac{1}{2}\epsilon^{12}\theta A_1\left(\partial_2A_2+F_{22}\right)-\frac{1}{2}\epsilon^{21}\theta A_2(\partial_1A_2+F_{12}),\\
        &=\frac{1}{2}Bx+\frac{1}{4}\theta Bx\left(\frac{1}{2}B+B)\right),\\
        &=\frac{1}{2}Bx+\frac{3}{8}\theta B^2x,\\
        &=\frac{1}{2}B\left(1+\frac{3}{4}B\theta\right)x.
    \end{aligned}
\end{equation}

Finally, one can calculate the noncommutative Field Strength using equation (5) of the paper, which is
\begin{equation}\label{eq:F_ij}
    \hat{F}_{ij}=\partial_i\hat{A}_j-\partial_j\hat{A}_i-i[\hat{A}_i,\hat{A}_j]_*.
\end{equation}
Using (\ref{eq:A_hat_x}) and (\ref{eq:A_hat_y} in (\ref{eq:F_ij}), one can calculate the field strength $\hat{B}$
\begin{equation}\label{eq:nc_B}
    \begin{aligned}
        \hat{B}_{\hbox{\tiny sym}}=\hat{F}_{12}&=\partial_1\hat{A}_2-\partial_2\hat{A}_1-i[\hat{A}_1,\hat{A}_2]_*,\\
        &=\frac{1}{2}B\left(1+\frac{3}{4}B\theta\right)+\frac{1}{2}B\left(1+\frac{3}{4}B\theta\right)+i\left[\frac{1}{2}B\left(1+\frac{3}{4}B\theta\right)\right]^2(-i\theta),\\
        &= B\left(1+\frac{3}{4}B\theta\right)+\theta\left[\frac{1}{2}B\left(1+\frac{3}{4}B\theta\right)\right]^2.
    \end{aligned}
\end{equation}
Let us define
\begin{equation}
    \overline{B}_{\hbox{\tiny sym}}=B\left(1+\frac{3}{4}B\theta\right).
\end{equation}
Therefore, equation (\ref{eq:nc_B}) is given by
\begin{equation}\label{eq:nc_B_bar}
    \begin{aligned}
        \hat{B}_{\hbox{\tiny sym}} =\overline{B}_{\hbox{\tiny sym}}+\frac{1}{4}\overline{B}_{\hbox{\tiny sym}}^2\theta.
    \end{aligned}
\end{equation}
From (\ref{eq:nc_B_bar}), one can express $\hat{B}$ in first order of $\theta$. The exact expression is given by
\begin{equation}
    \begin{aligned}
        \hat{B}_{\hbox{\tiny sym}}&= B\left(1+\frac{3}{4}B\theta\right)+\frac{1}{4}B^2\theta+\mathcal{O}(\theta^2),\\
        &= B+\frac{3}{4}B^2\theta+\frac{1}{4}B^2\theta+\mathcal{O}(\theta^2),\\
        &=B(1+B\theta)+\mathcal{O}(\theta^2).
    \end{aligned}
\end{equation}
From (\ref{eq:A_hat_x}) and (\ref{eq:A_hat_y}), It is evident that there is a contribution of noncommutative parameter $\theta$ in the noncommutative potential.

In the Landau gauge, the potential is given by
\begin{equation}
    A_1=-By \quad\text{and}\quad A_2=0.
\end{equation}
Again, one uses (\ref{eq:sw}) to calculate the NC potential $\hat{A}_1$ and $\hat{A}_2$. In particular, 
\begin{equation}
    \begin{aligned}
        \hat{A}_{1,\hbox{\tiny Lan}}&=A_1-\frac{1}{2}\epsilon^{12}\theta A_1(\partial_2A_1+F_{21})-\frac{1}{2}\epsilon^{21}\theta A_1(\partial_1A_1+F_{11}),\\
        &= -By+\frac{1}{2}\theta By(-B-B),\\
        &= -By-B^2\theta y,\\
        &= -B(1+B\theta)y,
    \end{aligned}
\end{equation}
and 
\begin{equation}
    \hat{A}_{2,\hbox{\tiny Lan}}=0.
\end{equation}
Furthermore, in the Landau gauge the NC field Strength is given by
\begin{equation}
    \hat{B}_{\hbox{\tiny Lan}}= B(1+B\theta).
\end{equation}
One uses the expression of $\hat{A}_x$ and $\hat{A}_y$ to construct the noncommutative superpotential according to equation (21) of the paper. The superpotential equation is
\begin{equation}
    \begin{aligned}
        \mathcal{A}= \frac{1}{\sqrt{2}}[i(\partial_x+i\hat{A}_x)+(\partial_y+i\hat{A}_y)].
    \end{aligned}
\end{equation}
The supercharge $Q$ is given by
\begin{equation}
    Q=\begin{pmatrix}
        0 & \mathcal{A}\\
        0 & 0
    \end{pmatrix},
\end{equation}
and the super Hamiltonian is the anti-commutator of $Q$ and $Q^\dag$, i.e.,
\begin{equation}
    H=\{Q,Q^\dag\}_*=\begin{pmatrix}
        \mathcal{A}\mathcal{A}^\dag & 0\\
        0 & \mathcal{A}^\dag\mathcal{A}
    \end{pmatrix}.
\end{equation}
In other words, the partner Hamiltonians are $H_1=\mathcal{A}^\dag\mathcal{A}$ and $H_2=\mathcal{A}\mathcal{A}^\dag$. The Hamiltonian in terms of the NC variable is given by
\begin{equation}\label{eq:susy_hamiltonian}
    \hat{H}=\frac{1}{2}\left[-(\partial_i+i\hat{A}_i)^2\mathbb{I}_{2\times 2}+\hat{B}\sigma_3\right].
\end{equation}
The equation above is the well-known Pauli Hamiltonian when the gyromagnetic ratio $g=2$.

To calculate the Eigenvalue of the Hamiltonian, let us start with the basics of the Harmonic oscillator Hamiltonian. The Hamiltonian of a simple Harmonic oscillator is given by
\begin{equation}\label{eq:H_oscilator}
    \begin{aligned}
        \hat{H}&=\frac{\hat{p}^2}{2m}+\frac{1}{2}m\omega^2\hat{x}^2,\\
        &= -\frac{\hbar^2}{2m}\frac{\mathrm{d}^2}{\mathrm{d}x^2}+\frac{1}{2}m\omega^2\hat{x}^2.
    \end{aligned}
\end{equation}
The above Hamiltonian can be solved using the algebraic method. Where the operator $\hat{a}$ and $\hat{a}^\dag$ is given by
\begin{equation}
    \begin{aligned}
        \hat{a}&= \sqrt{\frac{m\omega}{2\hbar}}\left(\hat{x}+\frac{i\hat{p}}{m\omega}\right),\\
        \hat{a}^\dag&= \sqrt{\frac{m\omega}{2\hbar}}\left(\hat{x}-\frac{i\hat{p}}{m\omega}\right).
    \end{aligned}
\end{equation}
One can easily show that the Hamiltonian is given by
\begin{equation}
    \hat{H}=\hbar\omega\left(\hat{a}^\dag \hat{a}+\frac{1}{2}\right).
\end{equation}
Whereas the eigenvalue is given by
\begin{equation}\label{eq:eigenvalue-HO}
    E_n=\hbar\omega\left(n+\frac{1}{2}\right).
\end{equation}
Where $n=0,1,2,\dots$. Now, let us set $\hbar=m=1$. Then equation (\ref{eq:H_oscilator}) can be expressed as
\begin{equation}
    \hat{H}= -\frac{\mathrm{d}^2}{\mathrm{d}x^2}+\frac{1}{2}\omega^2\hat{x}^2.
\end{equation}
One needs to change the operator $\hat{a}$ and $\hat{a}^\dag$ accordingly, i.e.,
\begin{equation}\label{eq:ladder}
    \begin{aligned}
        \hat{a}&= \sqrt{\frac{\omega}{2}}\left(\hat{x}+\frac{i\hat{p}}{\omega}\right),\\
        \hat{a}^\dag&= \sqrt{\frac{\omega}{2}}\left(\hat{x}-\frac{i\hat{p}}{\omega}\right).
    \end{aligned}
\end{equation}
Therefore, the Hamiltonian is given by
\begin{equation}
    \hat{H}=\omega\left(\hat{a}^\dag\hat{a}+\frac{1}{2}\right),
\end{equation}
and the energy
\begin{equation}
    E_n=\omega\left(n+\frac{1}{2}\right).
\end{equation}
According to (\ref{eq:susy_hamiltonian}), the supersymmetric Hamiltonian in the Landau gauge takes the form
\begin{equation}\label{eq:landau-Hamiltonian}
    \begin{aligned}
        \hat{H}_{\hbox{\tiny{Lan}}}&=\frac{1}{2}\left[(-\left(\partial_x-iB(1+B\theta)y\right)^2-\partial_y^2)\hat{\mathbb{I}}_{2\times 2}+B(1+B\theta)\sigma_{3}\right],\\
        &=\frac{1}{2}\left[(-\partial_x^2-\partial_y^2+2i\hat{B}_{\hbox{\tiny Lan}}y\partial_x+\hat{B}_{\hbox{\tiny Lan}}^2y^2)\hat{\mathbb{I}}_{2\times 2}+\hat{B}_{\hbox{\tiny Lan}}\sigma_{3}\right].
    \end{aligned}
\end{equation}
Here, the operator $\partial_x$ commutes with the Hamiltonian $\hat{H}_{\hbox{\tiny{Lan}}}$ expressed in the Landau gauge. Hence, one can replace the operator $\partial_x$ with its eigenvalue $ik_x$ so that the expression for $\hat{H}_{\hbox{\tiny{Lan}}}$ in (\ref{eq:landau-Hamiltonian}) can be manipulated as
\begin{equation}
    \begin{aligned}
        \hat{H}_{\hbox{\tiny{Lan}}}&= \frac{1}{2}\left[(-\partial_x^2-\partial_y^2-2\hat{B}_{\hbox{\tiny Lan}}yk_x+\hat{B}_{\hbox{\tiny Lan}}^2y^2)\hat{\mathbb{I}}_{2\times 2}+\hat{B}_{\hbox{\tiny Lan}}\sigma_{3}\right],\\
        &=\frac{1}{2}\left[(-\partial_x^2-\partial_y^2-2\hat{B}_{\hbox{\tiny Lan}}yk_x+\hat{B}_{\hbox{\tiny Lan}}^2y^2)\hat{\mathbb{I}}_{2\times 2}+\hat{B}_{\hbox{\tiny Lan}}\sigma_{3}\right],\\
        &=\frac{1}{2}\left[\left(-\partial_x^2-\partial_y^2+\hat{B}_{\hbox{\tiny Lan}}^2\left(y^2-2y\frac{k_x}{\hat{B}_{\hbox{\tiny Lan}}}+\frac{k_x^2}{\hat{B}_{\hbox{\tiny Lan}}^2}\right)-k_x^2\right)\hat{\mathbb{I}}_{2\times 2}+\hat{B}_{\hbox{\tiny Lan}}\sigma_{3}\right],\\
        &= \frac{1}{2}\left[\left(-\partial_x^2-\partial_y^2+\hat{B}_{\hbox{\tiny Lan}}^2\left(y-\frac{k_x}{\hat{B}_{\hbox{\tiny Lan}}}\right)^2-k_x^2\right)\hat{\mathbb{I}}_{2\times 2}+\hat{B}_{\hbox{\tiny Lan}}\sigma_{3}\right].
    \end{aligned}
\end{equation}
One can now compute the energy eigenvalues of the Hamiltonian $\hat{H}_{\hbox{\tiny{Lan}}}$ expressed in the Landau gauge following equation (\ref{eq:eigenvalue-HO}). The energy eigenvalues are given by
\begin{equation}\label{eq:energy-Landau-gauge}
    \begin{aligned}
        E^{\hbox{\tiny{Lan}}}_n&=\hat{B}_{\hbox{\tiny Lan}}\left(n+\frac{1}{2}\right)+\frac{\hat{B}_{\hbox{\tiny Lan}}}{2}\sigma,\\
        &=\hat{B}_{\hbox{\tiny Lan}}\left(n+\frac{1}{2}+\frac{1}{2}\sigma\right),\\
        &=B(1+B\theta)\left(n+\frac{1}{2}+\frac{1}{2}\sigma\right).
    \end{aligned}    
\end{equation}
One immediately observes from (\ref{eq:energy-Landau-gauge}) that
the ground state energy in the Landau gauge is zero as expected, i.e.,
\begin{equation}\label{eq:ground-state-energy-Landau-gauge}
E^{\hbox{\tiny{Lan}}}_{0}=0.
\end{equation}
It is important to note here that the eigenvalues of $\partial_x^2$ cancel out $-k_x^2$. Besides, there is no effect of the constant term $\frac{k_x}{\hat{B}_{\hbox{\tiny Lan}}}$ in the energy eigenvalue. In particular, the Harmonic oscillator potential is shifted in the coordinate space by an amount of $\frac{k_x}{\hat{B}_{\hbox{\tiny Lan}}}$. One should also note here that the energy of the Harmonic oscillator is not affected by the translation of the momentum.

Finally, the SUSY Hamiltonian in the symmetric gauge takes the form
\begin{equation}\label{eq:SUSY-Hamiltonian_Symmetric}
    \begin{aligned}
        \hat{H}_{\hbox{\tiny{Sym}}}= \frac{1}{2}&\Bigg[\left\{-\left(\partial_x-\frac{i}{2}B\left(1+\frac{3}{4}B\theta\right)y\right)^2-\left(\partial_y+\frac{i}{2}B\left(1+\frac{3}{4}B\theta\right)x\right)^2\right\}\hat{\mathbb{I}}_{2\times 2}\\
        &+B(1+B\theta)\sigma_{3}\Bigg],
    \end{aligned}
\end{equation}
which, upon simplification, leads to
\begin{equation}
    \begin{aligned}
        \hat{H}_{\hbox{\tiny{Sym}}} = \frac{1}{2}&\Bigg[\left\{-\partial_x^2-\partial_y^2+\frac{B^2}{4}\left(1+\frac{3}{4}B\theta\right)^2(x^2+y^2)+B\left(1+\frac{3}{4}B\theta\right)L_z\right\}\hat{\mathbb{I}}_{2\times 2}\\
        &+B(1+B\theta)\sigma_{3}\Bigg].
    \end{aligned}
\end{equation}
Using standard techniques, one can show that the energy eigenvalues of the Hamiltonian above expressed in symmetric gauge is given by
\begin{equation}\label{eq:eigenvalues-symmetric-Harikumaretal}
    E^{\hbox{\tiny{Sym}}}_{n,m}=B\left(1+\frac{3}{4}B\theta\right)\left(n+\frac{1}{2}+m+|m|\right)+\frac{1}{2}B(1+B\theta)\sigma.
\end{equation}
Here, $m\in\mathbb{Z}$ is the eigenvalue of the $L_z$ operator. Also, $n=0,1,2,...$. According to the above equation, it is evident that the ground state energy in the symmetric gauge is
\begin{equation}\label{eq:ground-state-energy-sym-gauge}
    \begin{aligned}
        E^{\hbox{\tiny{Sym}}}_{0,m}&=\frac{B}{2}\left(1+\frac{3}{4}B\theta\right)-\frac{1}{2}B(1+B\theta),\\
        &=\frac{3}{8}B^2\theta-\frac{1}{2}B^2\theta,\\
        &=-\frac{1}{8}B^2\theta,
    \end{aligned}
\end{equation}
for $m$ being zero or any negative integer. Now, based on (\ref{eq:ground-state-energy-Landau-gauge}) with (\ref{eq:ground-state-energy-sym-gauge}), one concludes that the ground state energy of the supersymmetric Hamiltonian given in (\ref{eq:susy_hamiltonian}) is not unambiguously zero as opposed to the claim in \cite{Harikumaretal}. And hence, it can not be concluded using the analysis undertaken in \cite{Harikumaretal} that the supersymmetry remains unbroken in this situation.

\subsection{Detailed Calculation associated with the article \texorpdfstring{\cite{Ashokdasetal}}{Lg}}\label{susec:appendix-ashokdasetal}
The NC supersymmetric Hamiltonian is given by the equation
\begin{equation}
    2H^{\hbox{\tiny (NC)}}=\{Q,Q^\dag\}_\star,
\end{equation}
where the anticommutator is taken with respect to the star product (see \cite{Ashokdasetal}, page 5). The supercharges in the above equation are given by
\begin{equation}\label{eq:D_supercharges}
    \begin{aligned}
        Q&=((\partial_1-i\mathcal{A}_1)+i(\partial_2-i\mathcal{A}_2))\sigma_+,\\
        Q^\dag&=(-(\partial_1-i\mathcal{A}_1)+i(\partial_2-i\mathcal{A}_2))\sigma_-,
    \end{aligned}
\end{equation}
where $\sigma_{\pm}=\frac{\sigma_1\pm\sigma_2}{2}$. These supercharges are $2\times 2$ matrices and hence it is evident that the supersymmetric Hamiltonian is a $2\times 2$ matrix.

Using the above expression of $Q$ and $Q^\dag$, one can simplify the NC supersymmetric Hamiltonian as
\begin{equation}
    2H^{\hbox{\tiny (NC)}}= (-(\partial_1-i\mathcal{A}_1)_\star^2-(\partial_2-i\mathcal{A}_2)_\star^2)\mathbb{I}_2+(\partial_1\mathcal{A}_2-\partial_2\mathcal{A}_1-i[\mathcal{A}_1,\mathcal{A}_2]_\star)\sigma_3,
\end{equation}
where $\mathbb{I}_2$ and $\sigma_3$ are the $2\times 2$ identity matrix and third Pauli matrix respectively. In the NC setup, one identifies
\begin{equation}\label{eq:D_NC_field_strength}
    \mathcal{F}_{12}=\partial_1\mathcal{A}_2-\partial_2\mathcal{A}_1-i[\mathcal{A}_1,\mathcal{A}_2]_\star,
\end{equation}
which is the generalization of the field strength tensor associated with the commutative gauge field.

In the paper, the authors considered the SUSY Landau problem in the NC space. In the commutative space, the vector potentials in the symmetric gauge are given by
\begin{equation}\label{eq:D_symmetric_gauge}
    \mathcal{A}_1=-\frac{B}{2}\hat{x}^2,\qquad \mathcal{A}_2=\frac{B}{2}\hat{x}^1,
\end{equation}
where $B$ represents the constant magnetic field and $\hat{x}_1,\hat{x}_2$ represents the quantum mechanical position operator. Here, one should note that we use a hat to represent quantum mechanical operator, i.e., observables, and without a hat the NC observables. Using (\ref{eq:D_supercharges}), the supercharges in terms of the vector potentials can be written as
\begin{equation}
    \begin{aligned}
        Q&=i\left(p_1+ip_2-\frac{iB}{2}(x^1+ix^2)\right)\sigma_+,\\
        Q^\dag &=-i\left(p_1-ip_2+\frac{iB}{2}(x^1-ix^2)\right)\sigma_-,
    \end{aligned}
\end{equation}
The supersymmetric Hamiltonian in the NC space is given by
\begin{equation}
    2H_{\hbox{\tiny sym}}^{\hbox{\tiny NC}}=\left(\boldsymbol{p}\star\boldsymbol{p}+\left(\frac{B}{2}\right)^2\boldsymbol{x}\star\boldsymbol{x}-BL\right)\mathbb{I}_2+F_{12}\sigma_3,
\end{equation}
where $\boldsymbol{p}=(p_1,p_2)$, $\boldsymbol{x}=(x^1,x^2)$, and $L=x^1p_2-x^2p_1$ is the third component of the orbital angular momentum in the noncommutative space.  Besides, one uses (\ref{eq:D_symmetric_gauge}) in (\ref{eq:D_NC_field_strength}) to calculate $F_{12}$. In particular,
\begin{equation}
    \begin{aligned}
        F_{12}&=\partial_1\left(\frac{B}{2}x^1\right)-\partial_2\left(-\frac{B}{2}x^2\right)-i\left[-\frac{B}{2}x^2,\frac{B}{2}x^1\right]_\star,\\
        &=B+i\left(\frac{B}{2}\right)^2[x^2,x^1]_\star,\\
        &=B+\frac{B^2}{4}\theta,\\
        &=B\left(1+\frac{1}{4}B\theta\right),
    \end{aligned}
\end{equation}
where one uses noncommutative commutation relation $[x^1,x^2]_\star=i\theta$. In the paper, the authors introduced Bopp shift (see \cite{Ashokdasetal}, page 1). To achieve the noncommutative commutation relation, the Bopp shift takes the form
\begin{equation}
    x^i\to \hat{x}^i-\frac{\theta}{2}\epsilon^{ij}\hat{p}_j.
\end{equation}
Note that under the Bopp shift, one can calculate
\begin{equation}
    \begin{aligned}
    p_1\star p_1\star\phi&=(\hat{p}_1)^2\phi.\\
    x^1\star x^1\star\phi&=\left(\hat{x}^1-\frac{\theta}{2}\hat{p}_2\right)\left(\hat{x}^1-\frac{\theta}{2}\hat{p}_2\right)\phi,\\
    &=\left((\hat{x}^1)^2+\frac{\theta^2}{4}(\hat{p}_2)^2-\frac{\theta}{2}\hat{x}^1\hat{p}_2-\frac{\theta}{2}\hat{x}^1\hat{p}_2\right)\phi,\\
    &= \left((\hat{x}^1)^2+\frac{\theta^2}{4}(\hat{p}_2)^2-\theta \hat{x}^1\hat{p}_2\right)\phi.\\
    x^2\star x^2\star\phi&=\left(\hat{x}^2+\frac{\theta}{2}\hat{p}_1\right)\left(\hat{x}^2+\frac{\theta}{2}\hat{p}_1\right)\phi,\\
    &=\left((\hat{x^2})^2+\frac{\theta^2}{4}(\hat{p}_1)^2+\frac{\theta}{2}\hat{x}^2\hat{p}_1+\frac{\theta}{2}\hat{x}_2\hat{p}_1\right)\phi,\\
    &=\left((\hat{x}^2)^2+\frac{\theta^2}{4}(\hat{p}_1)^2+\theta \hat{x}^2\hat{p}_1\right)\phi.\\
    \boldsymbol{L}\star\phi&=(x^1\star p_2-x^2\star p_1)\phi,\\
    &= \left(\left(\hat{x}^1-\frac{\theta}{2}\hat{p}_2\right)\hat{p}_2-\left(\hat{x}^2+\frac{\theta}{2}\hat{p}_1\right)\hat{p}_1\right)\phi,\\
    &=\left(\hat{L}-\frac{\theta}{2}\left((\hat{p}_1)^2+(\hat{p}_2)^2\right)\right)\phi.
    \end{aligned}
\end{equation}
Therefore, one can write
\begin{equation}
    \begin{aligned}
        \boldsymbol{p}^2\star\phi &=\left((\hat{p}_1)^2+(\hat{p}_2)^2\right)\phi.\\
        \boldsymbol{x}^2\star\phi &= \left((\hat{x}^1)^2+(\hat{x}^2)^2+\frac{\theta^2}{4}\left((\hat{p}_1)^2+(\hat{p}_2)^2\right)-\theta\hat{L}\right)\phi.\\
        \boldsymbol{L}\star\phi&=\left(\hat{L}-\frac{\theta}{2}\left((\hat{p}_1)^2+(\hat{p}_2)^2\right)\right)\phi.
    \end{aligned}
\end{equation}
Hence, the supersymmetric NC Hamiltonian is given by
\begin{equation}\label{eq:D_Hamiltonian_symmetric}
    \begin{aligned}
        2H_{\hbox{\tiny sym}}^{\hbox{\tiny NC}}&=\left(\hat{\boldsymbol{p}}^2\left(1+\frac{\theta B}{4}\right)^2+\left(\frac{B}{2}\right)^2\hat{\boldsymbol{x}}^2-B\left(1+\frac{\theta B}{4}\right)\hat{\boldsymbol{L}}\right)\mathbb{I}_2+B\left(1+\frac{\theta B}{4}\right)\sigma_3,\\
        &=\left(\hat{\overline{\boldsymbol{p}}}^2+\left(\frac{\mathcal{B}}{2}\right)^2\hat{\overline{\boldsymbol{x}}}^2-\mathcal{B}\hat{\boldsymbol{L}}\right)\mathbb{I}_2+\mathcal{B}\sigma_3,
    \end{aligned}
\end{equation}
where the following identifications were made
\begin{equation}
    \begin{aligned}
        \hat{\overline{\boldsymbol{p}}}&=\hat{\boldsymbol{p}}\left(1+\frac{\theta B}{4}\right),\\
        \hat{\overline{\boldsymbol{x}}}&=\frac{\hat{\boldsymbol{x}}}{1+\frac{\theta B}{4}},\\
        \mathcal{B}&=B\left(1+\frac{\theta B}{4}\right).
    \end{aligned}
\end{equation}
The energy eigenvalues of the Hamiltonian given in the equation (\ref{eq:D_Hamiltonian_symmetric}) is
\begin{equation}\label{eq:symmetric-gauge-eigenvalue-naive-minimal-prescription}
    E_{n_-,n+}^{\hbox{\tiny sym}}=\frac{\mathcal{B}}{2}(2n_-+1)\pm \frac{\mathcal{B}}{2}.
\end{equation}
Let us consider the Landau gauge now. The vector potential in this gauge is given by
\begin{equation}\label{eq:landau-gauge-ashokdasetal}
    \mathcal{A}_1=-Bx^2, \qquad \mathcal{A}_2=0.
\end{equation}
In this gauge, the supercharges are
\begin{equation}
    \begin{aligned}
        Q&=((\partial_1+iBx^2)+i\partial_2)\sigma_+,\\
        Q^\dag&=(-(\partial_1+iBx^2)+i\partial_2)\sigma_-,
    \end{aligned}
\end{equation}
Therefore, the Hamiltonian is given by
\begin{equation}
    \begin{aligned}
        2H_{\hbox{\tiny Lan}}^{\hbox{\tiny NC}}&=\left(-(\partial_1+iBx^2)_\star^2-(\partial_2)^2\right)\mathbb{I}_2+F_{12}\sigma_3,\\
        &= \left(p_1\star p_1+p_2\star p_2+2Bx^2\star p_1+B^2x^2\star x^2\right)\mathbb{I}_2+F_{12}\sigma_3.
    \end{aligned}
\end{equation}
In the Landau gauge, the field strength is given by
\begin{equation}
    F_{12}=B.
\end{equation}
Note that under the Bopp shift, one can calculate
\begin{equation}
    \begin{aligned}
        \boldsymbol{p}\star\boldsymbol{p}\star\psi&=\hat{\boldsymbol{p}}^2\psi.\\
        x^2\star p_1\star \psi&=\left(\hat{x}^2+\frac{\theta}{2}\hat{p}_1\right)\hat{p}_1\psi,\\
        &= \left(\hat{x}^2\hat{p}_1+\frac{\theta}{2}\hat{p}_1^2\right)\psi.\\
        x^2\star x^2\star\phi&=\left(\hat{x}^2+\frac{\theta}{2}\hat{p}_1\right)\left(\hat{x}^2+\frac{\theta}{2}\hat{p}_1\right)\phi,\\
        &=\left((\hat{x^2})^2+\frac{\theta^2}{4}(\hat{p}_1)^2+\frac{\theta}{2}\hat{x}^2\hat{p}_1+\frac{\theta}{2}\hat{x}_2\hat{p}_1\right)\phi,\\
        &=\left((\hat{x}^2)^2+\frac{\theta^2}{4}(\hat{p}_1)^2+\theta \hat{x}^2\hat{p}_1\right)\phi.\\
    \end{aligned}
\end{equation}
Now, using the above relation one can rewrite the supersymmetric Hamiltonian of the NC space as
\begin{equation}
    \begin{aligned}
        2H_{\hbox{\tiny Lan}}^{\hbox{\tiny NC}}&=\left[\hat{p}_1^2+2B\left(\hat{x}^2\hat{p}_1+\frac{\theta}{2}\hat{p}_1^2\right)+B^2\left((\hat{x}^2)^2+\frac{\theta^2}{4}\hat{p}_1^2+\theta\hat{x}^2\hat{p}_1\right)+\hat{p}_2^2\right]\mathbb{I}_2+B\sigma_3,\\
        &= \left[\hat{p}_1^2\left(1+B\theta+\frac{B^2\theta^2}{4}\right)+\hat{p}_2^2+B^2\left\{(\hat{x}^2)^2+2\frac{1}{B}\hat{x}^2\hat{p}_1\left(1+\frac{B\theta}{2}\right)\right\}\right]\mathbb{I}_2+B\sigma_3.
    \end{aligned}
\end{equation}
The operator $\hat{p}_1$ commutes with the above Hamiltonian, since there is no $\hat{x}^1$ in the Hamiltonian. Hence, they have the same eigenstates. Therefore, one can replace $\hat{p}_1$ with the eigenvalue $k_1$ of $\hat{p}_1$. The Hamiltonian can be written as
\begin{equation}
    \begin{aligned}
        2H_{\hbox{\tiny Lan}}^{\hbox{\tiny NC}}&= \Bigg[k_1^2\left(1+\frac{B\theta}{2}\right)^2+\hat{p}_2^2+B^2\left\{(\hat{x}^2)^2+2\frac{1}{B}\hat{x}^2k_1\left(1+\frac{B\theta}{2}\right)+\left(\frac{k_1}{B}\left(1+\frac{B\theta}{2}\right)\right)^2\right\}-\\
        &\qquad k_1^2\left(1+\frac{B\theta}{2}\right)^2\Bigg]\mathbb{I}_2+B\sigma_3,\\
        &= \left[\hat{p}_2^2+B^2\left\{\hat{x}^2+\frac{k_1}{B}\left(1+\frac{B\theta}{2}\right)\right\}^2\right]\mathbb{I}_2+B\sigma_3.
    \end{aligned}
\end{equation}
The expression in the square bracket is an expression of a 1-dimensional Harmonic oscillator. The energy eigenvalue of the Hamiltonian is given by the following expression
\begin{equation}\label{eq:eigenvalues-landau-gauge-SUSY-Hamiltonian-Naive-Minimal-prescription}
    E_n^{\hbox{\tiny Lan}}=B(n+\frac{1}{2}\pm\frac{1}{2}).
\end{equation}
Here, one should note that the eigenvalue is independent of the noncommutative parameter $\theta$. Besides, it is evident from the above expression of energy eigenvalue that the eigenvalue is independent of $k_1\in \mathbb{R}$. Hence, one can conclude that the eigenvalue is continuously degenerate.

\subsection{Detailed Calculation associated with the Ground State of the Fermionic Hamiltonian}\label{appendix:ground_state_of_fermionic_hamiltonian}

Since the ground state energy is zero for the Hamiltonian $H_1$, it is evident that the superpotential should annihilate the ground state. Therefore, one can write
\begin{equation}\label{eq:ground_state}
    \mathcal{A}*^r\psi_{0}^{(1)} = 0.
\end{equation}
Using (\ref{eq:super_potential_product}), one can write the left side of the above equation as
\begin{equation}\label{eq:A*r}
    \begin{aligned}
        \mathcal{A}*^r \psi_{0}^{(1)}&= i\Bigg[p_x*\psi_{0}^{(1)}+\frac{2(1-r)e\hbar B}{\hbar +\sqrt{\hbar^2-4r(r-1)e\hbar\vartheta B}}y*^r\psi_{0}^{(1)}\Bigg]\\
        &\;\;+\Bigg[-p_y*\psi_{0}^{(1)}+\frac{2re\hbar B}{\hbar +\sqrt{\hbar^2-4r(r-1)e\hbar\vartheta B}}x*^r\psi_{0}^{(1)}\Bigg],\\
        &= i\Bigg[\hat{p}_x\psi_{0}^{(1)}+\frac{2(1-r)e\hbar B}{\hbar +\sqrt{\hbar^2-4r(r-1)e\hbar\vartheta B}}\hat{Y}^r\psi_{0}^{(1)}\Bigg]\\
        &\;\;+\Bigg[-\hat{p}_y\psi_{0}^{(1)}+\frac{2re\hbar B}{\hbar +\sqrt{\hbar^2-4r(r-1)e\hbar\vartheta B}}\hat{X}^r\psi_{0}^{(1)}\Bigg],\\
        &= i\Bigg[\hat{p}_x\psi_{0}^{(1)}+\frac{2(1-r)e\hbar B}{\hbar +\sqrt{\hbar^2-4r(r-1)e\hbar\vartheta B}}\Big(\hat{y}+\frac{r\vartheta}{\hbar}\hat{p}_x\Big)\psi_{0}^{(1)}\Bigg]\\
        &\;\;+\Bigg[-\hat{p}_y\psi_{0}^{(1)}+\frac{2re\hbar B}{\hbar +\sqrt{\hbar^2-4r(r-1)e\hbar\vartheta B}}\Big\{\hat{x}+\frac{(r-1)\vartheta}{\hbar}\hat{p}_y\Big\}\psi_{0}^{(1)}\Bigg],\\
        &= i\Bigg[\frac{2(1-r)e\hbar B}{\hbar +\sqrt{\hbar^2-4r(r-1)e\hbar\vartheta B}}\hat{y}+\Bigg\{1+\frac{2r(1-r)e\vartheta B}{\hbar +\sqrt{\hbar^2-4r(r-1)e\hbar\vartheta B}}\Bigg\}\hat{p}_x\Bigg]\psi_{0}^{(1)}\\
        &\;\;+\Bigg[\frac{2re\hbar B}{\hbar +\sqrt{\hbar^2-4r(r-1)e\hbar\vartheta B}}\hat{x}-\Bigg\{1+\frac{2r(1-r)e\vartheta B}{\hbar +\sqrt{\hbar^2-4r(r-1)e\hbar\vartheta B}}\Bigg\}\hat{p}_y\Bigg]\psi_{0}^{(1)}.
    \end{aligned}
\end{equation}
For the sake of simplicity, let us choose
\begin{equation}\label{eq:MNQ}
    \begin{aligned}
        M&=\frac{2(1-r)e\hbar B}{\hbar +\sqrt{\hbar^2-4r(r-1)e\hbar\vartheta B}},\\
        N&= \frac{2re\hbar B}{\hbar +\sqrt{\hbar^2-4r(r-1)e\hbar\vartheta B}},\\
        S&=1+\frac{2r(1-r)e\vartheta B}{\hbar +\sqrt{\hbar^2-4r(r-1)e\hbar\vartheta B}}.
    \end{aligned}
\end{equation}
Using equation (\ref{eq:A*r}) and (\ref{eq:MNQ} in (\ref{eq:ground_state}), one can write
\begin{equation}\label{eq:ground_state_dif_eq}
    \begin{aligned}
        \Big[(iM\hat{y}+iS\hat{p}_x+N\hat{x}-S\hat{p}_y)\psi_{0}^{(1)}\Big](x,y)&=0,\\
        \Big(iMy+\hbar S\frac{\partial }{\partial x}+Nx+i\hbar S\frac{\partial}{\partial y}\Big)\psi_{0}^{(1)}(x,y)&=0.
    \end{aligned}
\end{equation}
The above first-order partial differential equation can be solved using the separation of variables technique. To this end, let us consider
\begin{equation}\label{eq:separation}
    \psi_{0}^{(1)}(x,y)=X(x)Y(y).
\end{equation}
Hence, equation (\ref{eq:ground_state_dif_eq}) takes the form
\begin{equation}
    \begin{aligned}
        iMy+\frac{\hbar S}{X}\frac{\partial X}{\partial x}+Nx+i\frac{\hbar S}{Y}\frac{\partial Y}{\partial y}=0.
    \end{aligned}
\end{equation}
One can write the above equation as
\begin{equation}
    \begin{aligned}
        \frac{\hbar S}{X}\diff{X}{x}+Nx&=m,\\
        iMy+i\frac{\hbar S}{Y}\diff{Y}{y}&=-m.
    \end{aligned}
\end{equation}
Here, $m$ can take any real value carrying the dimension of momentum. The solution of the above two first order linear differential equations is
\begin{equation}\label{eq:ground_state_XY}
    \begin{aligned}
        X(x)&= \exp\left[\frac{1}{\hbar S}\left(mx-\frac{Nx^2}{2}\right)+c\right],\\
        Y(y)&=\exp\left[\frac{1}{\hbar S}\left(imy-\frac{My^2}{2}\right)+c'\right].
    \end{aligned}
\end{equation}
Here, $c$ and $c'$ are dimensionless constants that can be determined using boundary conditions imposed on the wavefunction $\psi_{0}^{(1)}$ in (\ref{eq:separation}).

Using the above expression of $X(x)$ and $Y(y)$ and plugging in the values of $M,N$, and $S$, from (\ref{eq:MNQ}), one can write down the ground state wave function $\psi_{0}^{(1)}(x,y)$ for arbitrary real value of the gauge parameter $r$ explicitly as
\begin{equation}\label{eq:r-dependent-wavefunc}
    \begin{aligned}
        \psi_{0,r}^{(1)}(x,y)&= \exp\Bigg[\frac{1}{\hbar\left\{1+\frac{2r(1-r)e\vartheta B}{\hbar +\sqrt{\hbar^2-4r(r-1)e\hbar\vartheta B}}\right\}}\Big\{mx+imy-\frac{re\hbar Bx^2}{\hbar +\sqrt{\hbar^2-4r(r-1)e\hbar\vartheta B}}\\
        &\qquad\qquad-\frac{(1-r)e\hbar By^2}{\hbar +\sqrt{\hbar^2-4r(r-1)e\hbar\vartheta B}}\Big\}+k\Bigg].
    \end{aligned}
\end{equation}
Here, $k$, again, is a new dimensionless constant obtained from $c$ and $c^\prime$.

\end{document}